\documentclass[12pt]{article}
\usepackage{graphics}

\newcommand{\beq}{\begin{equation}}
\newcommand{\eeq}{\end{equation}}

\newcommand{\pdr}{\partial}
\newcommand{\beqs}{\begin{eqnarray}}
\newcommand{\eeqs}{\end{eqnarray}}

\newcommand{\half}{\frac{1}{2}}
\newcommand{\eps}{\epsilon}

\newcommand{\grad}{\nabla}
\begin{document}
\author{S. G. Rajeev
\\ Department of Physics and Astronomy\\
University of Rochester\\
Rochester,NY 14627 }

\def\m#1{{$#1$}}
\def\p{{ p}}
\def\q{{ q}}
\def\k{{ k}}

\def\hi{{\cal H}}
\def\ignore#1{{}}

\begin{center}
   {\LARGE\bf   Bound States in Models of Asymptotic Freedom }

   {\large\bf  S. G. Rajeev}
   \footnote{{\it Department of Physics and Astronomy, University of Rochester,
    Rochester, New York 14627,USA} \\}\\
   \vspace{.5cm}
   Institut Mittag-Leffler\\
   The Royal Swedish Academy of Sciences\\
   Aurav\"agen 17, Djursholm, Sweden\\
   \vspace{2cm}
   {\large\bf Abstract}
\end{center}

		We study some quantum theories which are divergent but
		for which renormalization can be performed
		nonperturbatively and explicitly. The result is a
		well--defined, finite formulation of these theories,
		in which neither a cutoff nor a bare coupling constant
		appears. Such theories describe `contact' interactions
		between particles which are encoded into the boundary
		conditions of the wavefunction rather than in the
		hamiltonian.  It is the attempt to describe them in
		terms of conventional potentials or self-interactions
		that lead to divergences.

		We discover that, after
		renormalization, the dynamics is described by  a new operator
		(the `principal operator' \m{\Phi(E)} ):
		the Schr\"odinger equation
		for energy levels is replaced by the eigenvalue
		problem \m{\Phi(E)\psi=0} for this operator. (More
		generally, the resolvent of the hamiltonian has an
		explicit formula in terms of the inverse of
		\m{\Phi(E)}.) Moreover
                the interactions are not described by  the boundary
		conditions (domain) of the operator,
		but by the formula for it; i.e., 
		by its action on smooth functions. Even when
		the theory cannot be explicitly solved, it can be
		given a finite formulation once this operator is
		determined. It is then possible to apply standard
		approximation methods; in particular we can determine the
		energy of bound states, which would be impossible in
		petrurbation theory.

		We propose to call such theories which are apparently
		divergent, but have a finite formulation in terms of
		the principal operator, {\em transfinite quantum
		theories}. We construct some examples of such
		transfinite quantum field theories:
		 quantum mechanics with `contact' interactions,
		three body problem with contact interactions, quantum
		fields (fermionic and bosonic) interacting with a
		point source, many body problems with contact
		interactions, and  non-relativistic field theory with
		polynomial interactions .
			
As an application we develop a theory of
self-interacting Bose fields in two dimensions  with an
attractive self-interaction. The ground state  is a
Bose-condensate for which 
the conventional many body theory breaks 
down due to divergences. The magnitude
of the ground state energy grows {\it exponentially} with the number of
particles, rather than like a power law as for conventional many body systems.

\begin{flushleft}
{\it PACS}: \\ {\it Keywords}:
\end{flushleft}
\pagebreak

\section{Introduction}

There is a fact about classical mechanics that is so basic that we
dont even call it a law of mechanics. Perhaps it should
be called the zeroth law of mechanics:

{\it All particles behave like free particles for short enough
intervals of time}.\hfill\break
In the language of modern analysis, the path of any particle is
differentiable, so that it can be approximated by a straight line for
small enough intervals of time.  There are other physical phenomena (such as diffusion) where this
law does not hold. Hence it is  not a self-evident
fact, but rather  a law of nature.  Newton's second law states  that the
deviation of the path from that of a  free particle 
is second order in time (when the path is
viewed as a curve in configuration space) and  is
given by the force divided by the mass. 

In quantum mechanics we have a similar zeroth law as well. The
propagator (i.e., the integral
kernel \m{h_{t}(x,y)=<x|e^{-Ht}|y>}, where \m{H=p^2+V(x)} is the hamiltonian
operator)    is a gaussian upto power law corrections:
\beq
	h_{t}(x,y)={e^{-{(x-y)^2\over 4t}}\over {(4\pi t)^{d\over 2}}
}\bigg(1+
				O(t)\bigg).
\eeq
(Of course this statement, just like the previous statement in classical
mechanics, is true only for non-singular potentials. The quantum theory
is more forgiving of singularities than the classical theroy. We will consider some singular
situations later.)

Now let us look at quantum field theory. The analogous statement would
be that the  correlation functions  \m{G(x,y)=<\phi(x)\phi(y)>} 
of a quantum field \m{\phi} will
approach those of a free field theory when the distance \m{|x-y|} 
 between points
becomes small. This is however {\it not } true for realistic quantum
field theories. In Quantum Electrodynamics for example, quite the
opposite is true: the smaller the distance between points the {\it larger}
the deviation of  a correlation function from that of the free theory.
Asymptotically free theories (such as Quantum ChromoDynamics) are
better behaved: their correlators do approach those of the free theory
{\it but} the deviations are of order \m{1\over \log |x-y|} rather than
\m{|x-y|}.

Thus if we apply conventional ideas of dynamics to such quantum field
theories we will run into divergences. These have to be removed by the
unwieldy procedure of renormalization. This leads us to ask if there
is another way to formulate such theories: to describe the dynamics of
the theory by some  operator (instead of  the hamiltonian) which is able to
describe the  logarithmic deviation  from the  free theory. We must first
construct such a new picture by working within  the traditional renormalization
method applied to simple model problems. Once we learn the basic ideas
we might be able to deal with more realistic situations. In this paper
we will propose such a reformulation of  renormalization by studying 
 a series of examples, starting from quantum mechanics
(with singular potentials) and ending with some non-relativistic field
theories.

 Such a new point of view is necessary, since the  most
  fundamental theory of nature-the standard model of elementaty
  particles-  is  a divergent quantum field theory.
   It  was discovered in part by the
 requirement of renormalizability: that  the divergences can be removed order by
order in perturbation theory by redefining the parameters. The main
remaining problems in the standard model are  due to the unpleasantness
left over from this procedure. For example, the `naturalness' problem
in the Higgs sector is an extreme sensitivity (quadratic dependence )
of observables (like the Higgs mass) on microscopic parameters.

There are also more practical reasons to look for  a new formulation
of renormalization. One of the main obstacles to solving for the dynamics of non-abelian
gauge theories  is also that its quantum theory is divergent. We only know
how to remove these divergences by perturbative renormalization. But
perturbation theory cannot describe the formation of bound states: a
serious problem since {\em all }  the observable states of the theory
are expected to be bound states (confinement). 

There are also motivations that come from outside of particle physics.
We don't yet have a quantum theory of gravity  due to the
divergences that arise in quantizing general relativity: this time the
infinities cannot be removed by renormalization. Perhaps a better
understanding of renormalization will lead to a way of quantizing
non-renormalizable theories as well: there might a be a non-trivial
fixed point for the renormalization group.

String theory is a serious candidate for a quantum theory of
gravitation and, possibly, even a unified theory of all forces. Moreover it
is a finite theory. Thus it may be  the ultimate solution for
many of the problems noted above. However, it should be possible to
account for the spectacular success of renormalizable quantum field
theories at current energies without having to resort to a new theory
of spacetime at Planck energies. There should be a consistent
nonperturbative formulation of quantum field theories whether or not
the ultimate theory is a finite string theory. To draw an analogy, it is
possible to have a mathematically consistent formulation of continuum (fluid)
mechanics even though the ultimate description of fluids is in terms
of a  finite (but enormously large ) number of particles. Quantum field theory
should have a self-contained formulation  whether or not it is just an
approximation to a more fundamental theory.

We also mention another  great success of the idea of renormalization:
the divergences that afflicted the theory of second order phase
transitions were finally understood by Wilson's development of the
renormalization group. New ideas on renormalization might also deepen
our understanding of phase transitions. We will give some examples in
this paper.

Thus, at the heart of each of the fundamental problems of modern theoretical
physics is a divergent quantum theory.
Progress is not possible without a better understanding  of these
infinities. The situation is analogous to knowing the basic laws of
mechanics but without knowing calculus: anything beyond the simplest examples
are inaccessible.  Indeed it was a deeper understanding of the nature
of infinity (Cantor's work on the transfinite numbers) that made
modern analysis possible and by extension the modern theory of
dynamical systems.  We must look for examples which will help us
develop such a general theory of divergent quantum systems: a truly
{\em analytical } quantum mechanics.

We will start with simple quantum mechanical models with singular
potentials \cite{renqm} and progress to systems with several degrees
 of freedom, the three-body problem and then  to
 non-relativistic field theories.  Eventualy we  hope to  formulate
QCD in this way, but even the cases we have
studied so far address some physically interesting problems.

To be more precise, we have found  a way to  reformulate some quantum theories that are divergent in
the usual formulation in a new
way that is manifestly finite: there is no need for a regularization
and all the parameters are physical, not `bare'  coupling constants.
We call such systems   {\em transfinite quantum systems}.
In these systems, interactions are not specified by potentials or any
other simple modification of the hamiltonian operator: indeed on smooth wavefunctions
the hamiltonian acts just like the free hamiltonian. It is the attempt to
shoehorn
these system into a conventional description in terms of a hamiltonian
such as
\m{H=-\nabla^2+V(x)} that leads us to divergences.
The interactions  are encoded not into the formula for the
hamiltonian but
 into the
boundary conditions  on the
wavefunctions (i.e., domain of self-adjointness of the hamiltonian).  Thus the
resolvent
of the hamiltonian (which is a sort of Green's function) contains the
complete specification of the system.   The resolvent
has the complete physical description of the system, in terms of
physical and not bare parameters.

Upto this point  the ideas are not very new: the resolvent  has been worked
out in many simple cases, most notably by Krein and his school \cite{albeverio}.
Unfortunately in most cases,
determining the resolvent is the same as exactly solving the
system. We need a way to think of such `contact'  interactions which
is free from divergences and yet does not require us to solve the
whole dynamics first: we just need to solve the `singular part' of the
dynamics. Our main discovery  is that there is a new
operator, which we call  the {\em Principal Operator} \m{\Phi(E)}, which
describes the dynamics of transfinite quantum systems. It is free of
divergences, can be
determined explicitly (usually as an integral operator) and is quite
simple in many cases. There are no subtleties in the definition of its
domain: the interaction is described  by a term in the formula for  
\m{\Phi(E)}.  In this way the formulation of transfinite quantum
systems is no more difficult than that of finite quantum systems.

The eigenvalues of energy are given by the
solutions to \m{\Phi(E)|\psi>=0}; the scattering amplitude is
determined by the inverse of \m{\Phi(E)}. Of course the solution of
the eigenvalue equation or the inversion of \m{\Phi(E)} is a difficult
dynamical problem: as difficult as solving the Schrodinger equations
would be in finite quantum systems. But the point is that it is no
more complicated than that: we can apply the standard methods such
as variational principles or perturbation theory to the principal
operator, since now we have a formulation free of divergences.
The system need not be exactly solvable for it to have a transfinite
formulation. Formulating the system  amounts to finding the principal operator
\m{\Phi(E)} while solving it amounts to inverting \m{\Phi(E)}.

The principal operator can be thought of as an effective hamiltonian
obtained after `integrating out' (or eliminating) the short distance
degrees of freedom. This is reminiscent of Wilson's program of
renormalization. The main new point is that we can get a closed
experession for such an operator, rather than give an implicit
prescription to find it as in Wilson's program.

Simple, exactly solvable , examples of such transfinite quantum
theories have been known for a long time: the two dimensional delta function
potential is a
good example \cite{huang}.  We will in fact start our discussion with
such  a  simple
example, the pole model of low energy scattering in quantum mechanics.
It has been known for a long time that  at momenta small compared to
the size of the scatterer, the scattering
amplitude of a particle tends to a universal form \m{f(k)={1\over \xi^{-1}-ik}}, the
number \m{\xi} being the {\em scattering length}. For positive
scattering length there is a bound state near threshold and for
negative scattering length a resonance (`virtual bound state').
But there is no finite quantum system that can give this simple model
of scattering amplitude: such a system can only be constructed as a
limit of  hamiltonians. Indeed this is a beautiful example of
the idea of renormalization at work. We will construct a transfinite
quantum theory of two body interactions that describes exactly this
case of low energy scattering.

These ideas are of   interest in modern (late 1990's) atomic physics.
Technical advances in the cooling and trapping of atoms have made it
possible very recently to  study experimentally the interaction of
atoms at
very low
momenta: i.e., wavelengths large compared to the size of the atoms
themselves. Scattering lengths for several species  have been measured.
In the case of Rubidium and Sodium the scattering length is positive
while for Lithium it is negative. (e.g., \m{\xi=-27.3 \pm 0.8} Bohr radii for
Lithium.) Interspecies scattering also should give a whole range of
values of scattering lengths, both positive and negative.
Indeed it is even possible to `tune' the scattering length to any desired value
using Feshbach resonances. Thus many of the phenomena
associated  to
low energy scattering have become or will soon become experimentally
accessible.  Of greatest interest is the formation of condensates of a large
number
of atoms.  This would be described by a non-relativistic field theory
(bosonic or  fermionic) with a contact interaction. The techniques
developed may eventually  be useful to study phase transitions in such
atomic condensates. So far we have a way to describe condensates in
two dimensional systems.

We will then pass to studying some field theoretical models.
T. D. Lee \cite{tdlee}  introduced a simple model for renormalization which has
been
studied further by many people. It describes a  Bose
field interacting with a point-like source. The source itself has two
possible states, which  can be thought of as describing the internal
states of  a heavy particle.  In the limit when the size of the source
goes to zero, there is an ultra-violet divergence in this model: for
example,
the energy difference between the two states of the source is
infinite. This can be removed by a renormalization procedure: the
parameters of the model are made to depend on a cut-off
( the  size of the source ) in
such a way that the  energies of physical states are finite in the
limit as the cut-off is removed.

An even simpler, non-relativistic, limit of the Lee model was studied
in ref. \cite{henleythirring} ). The heavy particle was interpreted as
the nucleon, its two states being the neutron and the proton. The
light (but  non-relativistic) boson was thought of as the charged
pion. The energy of the `neutron' (more precisely the neutron-proton mass
difference)  is
infinite.  This divergence can be regularized by requiring the
source to have a finite size \m{\Lambda^{-1}}-- an ultra-violet
cutoff.  Then we require a bare parameter  \m{\mu_\Lambda}
depend on \m{\Lambda} in such a way that in the limit we get a finite
answer for the neutron energy.

Now we have much better descriptions of the nucleon pion system. The
static source model is  thought of these days as a historical curiosity.
But this kind of model has, been  very
valuable as a  proving ground for   new approaches to
renormalization. The most
spectacular example has been the work of Wilson, who also perfected  his
ideas on a variant of the static source model \cite{kwilson}.  A
later (one dimensional) version of the static source  model, the Kondo
model, has become a classic example  of renormalization.
We will also use the static source model and (later on, the Kondo
problem) to test our ideas on
 renormalization. 

The simplest example of such a field theory we will study is a system
of (non-relativistic) fermions interacting with a point source.  This
fermionic variant of the Lee model also has a static source with two states:
one of these states is now a fermion and the other a boson. We will
show that the ultra-violet divergences can be removed by a
nonperturbative renormalization method. The energies are all finite
and the theory in fact has a compeltely finite description, with no
cutoffs or bare coupling constants in sight.
The main lesson we learn is that renormalizable quantum field 
theories of this type
(  `transfinite')  also have interactions that arise from
the boundary conditions of the wave function. Hence the dynamical
information is not encoded in the hamiltonian, but in the 
 the principal operator, \m{\ \Phi(E)}, which we construct. The spectrum of the
principal operator determines the energy levels: the Schr\"odinger
equation is replaced by the equation \m{\Phi(E)\psi=0}.  Although the problem
is not
exactly solvable, we can apply traditional approximation methods such
as the variational principle once we have  a finite form of the
theory.
We won't be able to get all the energy levels: but we will show that
they are all finite and determine the ground state and first excited
state energies.

We will then apply our methods to the bosonic static source model (the
original Lee model). This
case is more subtle.
In fact the  original analysis of the Lee model was incomplete: the
divergences were shown to be removed only in states which contain
at most two bosons. The possibility remained  that the hamiltonian of the
system is unbounded below, when the number of bosons is more than  two.
This  is not a mere  technicality: there
are systems in which an analogous  renormalization
model still leaves behind divergent energies for multi-particle
states  \cite{adhikariprl}. ( We will give such an example ourselves
later on in this  paper.) We will show that the non-relativistic
Lee model has  energies
bounded below in each sector with a fixed number of bosons ( this
number is a conserved quantity of this model.)  Moreover we will make
a variational estimate of the ground state energy of the Lee model in
the limit of a large number of bosons. Thus it will be established
that the Lee model is indeed free of divergences.

Next, we will apply these ideas  to the case of a system of
non-relativistic bosons in
two dimensions, interacting though a contact interaction:
non-relativistic \m{\lambda\phi^4} theory in \m{2+1} dimensions. 
This model has been studied in various guises by other authors as
well  \cite{Hoppe, Bergman,tarrach}, but our approach is somewhat different.
We
will obtain a closed form for the principal operator after
renormalization. We will then show how to solve the many body problem
in the mean field approximation. In fact we will obtain a solution for
the wavefunction of the bosonic condensate in this aproximation, as
well as its energy.

\section{Scattering at Low Momentum and Renormalization}

 In the
limit of small momentum we should expect that the scattering  amplitude
of two atoms does not depend on the
details of the atomic form-factors: the `shape' of the atoms should
not matter at wavelengths much larger than the atoms. This is a
situation to which the philosophy of renormalization applies
perfectly: the cut-off is the size of the atom: the independence of
the atomic scattering amplitude on the shape of the atom arises from
the independence on the regularization scheme of the renormalized
system. We will now show in some detail how to perform this renormalization.

Elastic scattering \cite{landau} of a  particle by a heavy target is described
by the scattering amplitude
\m{f(k,{\bf  n},{\bf  n}')}.  ( Of course
the case of two-parcticle scattering can be reduced to this case by
passing to the center of mass frame.) It is defined
in terms of the asymptotic
form of the wavefunction as \m{r\to \infty},
\beq
	\psi({\bf k},{\bf x})\sim e^{ikr{\bf n}\cdot {\bf
	n}'}+f(k,{\bf n},{\bf n}'){e^{ikr}\over r}.
\eeq
Here, \m{{\bf k}=k{\bf n}} is the momentum of the incoming wave
  and \m{{\bf n}'={{\bf x}\over r}} is  the direction of the outgoing wave.
Conservation of probability requires the ``unitarity condition'' on
the scattering amplitude:
\beq
	 f(k,{\bf n},{\bf n}')-f^*(k,{\bf n}',{\bf n})=
2ik\int f(k,{\bf n},{\bf n}'')f^*(k,{\bf n}', {\bf n}''){d\Omega_{{\bf
n}''}\over 4\pi}.
\eeq
Let us also recall the formula for the total scattering cross-section:
\beq
\sigma(k)=\int |f(k,{\bf n},{\bf n}')|^2 d\Omega_{{\bf n}'}.
\eeq

If a wave is scattered by a target whose size is small  compared
with the wavelength, only the partial wave with zero
angular  momentum will scatter: only this partial wave has  an
appreciable  probability of interacting with the target. Thus the
scattering  amplitude  will
become independent of angles in this limit, {\it even if the target is not
spherically symmetric}. The unitarity condition
 will  then require that
\beq
	{\rm Im}f(k)=k|f(k)|^2.
\eeq
In other words
\beq
	{\rm Im}{1\over f(k)}=-k
\eeq

 Thus, in the limit of vanishing  momentum the scattering amplitude
approaches a real constant (the unitarity condition requires the
imaginary part to be \m{O(k)}):

\beq
	\lim_{k\to 0}  f(k,{\bf  n},{\bf n}')=-\xi.
\eeq
The quantity  \m{\xi} is  called the ``scattering length''. It can be either
 positive or negative.
It is stated in some  textbooks  that a positive scattering length corresponds
to a
 generally repulsive interaction and a negative one to an attractive
 interaction. But this is only  true of interactions that are weak
 enough to be treated in the Born expansion. We will in fact see that a
 positive scattering length can even lead to a bound state.

Now the imaginary part of \m{1\over f(k)} must be  \m{O(k)} to satisfy
the unitary condition. In the limit of small but nonzero momentum we
thus have,
\beq
	{1\over f(k)}=-ik-{1\over \xi}
\eeq
or,
\beq
f(k,{\bf n},{\bf n}')=-{1\over \xi^{-1}+ik}.
\eeq

In fact  the scattering amplitude of {\it any}  target
whose  size is small compared to the wavelength  is
described asymptotically by this formula: even if the
target  has no special property such as spherical symmetry. We will
call this  the
 `simple pole model' for low
momentum scattering. (Due to Wigner(1933) and Bethe and Peierls
(1935).)   The scattering cross-section of
the target will be, in this model,
\beq
\sigma(k)=4\pi \xi^2{1\over 1+k^2\xi^2}.
\eeq
This gives the geometrical meaning of the scattering length: at low
momenta the target will appear to be a hard sphere of radius \m{\xi}.

The scattering amplitude has a simple pole at
\beq
	k={i\over \xi}.
\eeq
This pole is in the upper half of the complex \m{k}-plane  for
\m{\xi> 0}; it then  corresponds to a bound state of binding
energy  \m{1\over 2m\xi^2}.
If \m{\xi} is negative, the pole does not describe a  bound
state. Poles of \m{f(k)} in the lower half  of the \m{k}-plane
correspond to solutions of the Sch\"odinger equation that grow at
infinity: it corresponds to what is called a `virtual level'.( See
\cite{landau} section 133.)

This simple picture for the scattering  at low momenta however has  an
important  peculiarity: {\em there is no hamiltonian of the form
\beq
	H={{ p^2}\over 2m}+V({ x})
\eeq
with a potential function \m{V({ x})} which can reproduce such a
scattering amplitude exactly}. If there were  such a potential, it
 would have zero range
and infinite height (for positive \m{\xi}) or depth ( for negative
\m{\xi}.) \m{V({ x})} cannot  be described even by a familiar distribution such
as a delta function: the delta function potential in three dimensions
either has vanishing scattering amplitude (repulsive case) or has no
well-defined ground state (attractive case).

Instead we have to view
the hamiltonian   as arising from a {\it limit} of potentials, with  widths
tending  to zero; the heights have to be carefully  adjusted as a
function of  the
widths in this limit in order to get  a non-vanishing scattering
amplitude or a finite ground state energy. This  process of obtaining
a finite scattering amplitude from the limit of a sequence of
potentials with finite width is reminiscent  of the renormalization
program of quantum field theory; the width is the `short-distance
cut-off'  and the height (or depth) of the potential the `bare coupling
constant'. The main difference from conventional quantum field theory
is that the renormalization has to be  carried out non-perturbatively,
since we expect to recover a bound state in some cases. The answer
will be independent of the details of the limiting process.

We will now show that there is in fact a perfectly well--defined
quantum theory with the simple pole model above as the scattering
amplitude. Its hamiltonian is the same as that of the free particle
as far its action on smooth wavefunctions in position space is
concerned.
 The interactions are encoded into the boundary conditions
of the wavefunction
 at short distances ( or at infinity in momentum space). Thus the
hamiltonian is practically useless as a tool in studying this
system. We will instead obtain a  formula for the resolvent of the
regularized hamiltonian and take its limit as the cutoff is
removed. It will turn out that it is determined in terms of a function
\m{\phi(E)} (the `principal function'): we get a version of
the Krein formula for resolvents. The zeros of this principal function
give the point spectrum of the renomalized system.

When we have more than two particles, we can still preform the
renormalization as before and  get a formula for the
resolvent. However, this time the principal function is replaced by an
operator. In the case of the three body problem, we can study the
spectrum of this operator. If the dimension of space is two, we show
that this problem is well-posed and has a well--defined ground state
energy. In the case of three-dimensional three body problem, the
ground state energy of the renormalized theory still diverges: there
are further renormalizations necessary. Thus even if we can obtain a
Krein formula and a principal operator, we still need to show that the
spectrum is bounded below in order to have  a well--defined
theory. This is why we take pains to establish this lower bound in the case of
some  quantum field theories.

\subsection{Renormalized Resolvent}

Consider a pair of particles    with an
 attractive  short
range interaction.  After separating out the center of mass variable,
we can reduce this to the scattering of a particle of mass \m{m}
 (equal to the reduced mass of the pair)
against an immovable `target' representing the interaction between the
particles. We are interested in the limit as the
inverse range  of the interaction  \m{\Lambda} is very large compared to the
momentum of the particle.

This can be modelled by  the  Hamiltonian  operator, in momentum space,
\beq
H_\Lambda\psi(p)={\p^2\over 2m}\psi(\p)-
g(\Lambda)\rho_\Lambda(\p)\int \rho_{\Lambda}(\q)\psi(\q)[d\q].
\eeq
Here, \m{\rho_\Lambda(\p)} is a function that is equal to one near the
origin and falls off rapidly at infinity. For example,
\beq
	\rho_{\Lambda}(\p)=\theta(|{\bf p}|<\Lambda)
\eeq
would be a typical choice. The function \m{g(\Lambda)} will be picked
later in such a way that the scattering amplitude has a limit as
\m{\Lambda\to \infty}.

Here, the interaction is represented by a separable kernel which makes the
calculations simple. If  instead we choose a potential of range
\m{{1\over \Lambda}} we will get similar answers in the limit
\m{\Lambda\to \infty} but the calculations are more complicated.

Consider the inhomogenous equation
\beq
	({\p^2\over 2m}-E)\psi_\Lambda(\p)-
g(\Lambda)\rho_\Lambda(\p)\int \rho_\Lambda(\q)\psi_\Lambda(\q)[d\q]=\chi(\p).
\eeq
Then
\beq
	\psi_\Lambda(\p)={\chi(\p)\over {\p^2\over 2m}-E}+{A_\Lambda\over
{\p^2\over 2m}-E}\rho_\Lambda(\p)
\eeq
where \m{A_\Lambda} is
\beq
	A_\Lambda=g(\Lambda)\int \rho_\Lambda(\p)\psi_\Lambda(\p)[d\p].
\eeq
We can  put the expression  
 for \m{\psi_\Lambda(\q)} into this equation for \m{A_\Lambda}
to get,
\beq
A_\Lambda\bigg[g^{-1}(\Lambda)-
\int {\rho_{\Lambda}^2(\p)\over
{\p^2\over 2m}-E}[d\p]\bigg]=\int \rho_\Lambda(\p){\chi({\bf p})\over
{\p^2\over 2m}-E}[d\p]
\eeq
We now choose \m{g(\Lambda)} such that
\beq
	g^{-1}(\Lambda)-
\int {\rho_{\Lambda}^2(\p)\over
{\p^2\over 2m}-E}[d\p]
\eeq
has a limit as \m{\Lambda\to \infty}:
\beq
	g^{-1}(\Lambda)=\int \rho_{\Lambda}^2(\p){1\over
{\p^2\over 2m}+{\mu^2\over 2m}}[d\p]
\eeq
for some real constant \m{\mu}.

This number \m{\mu}( which has the dimension of momentum) is the true 
physical parameter which describes the
strength of the interaction: it remains meaningful even in the limit
as \m{\Lambda} goes to infinity. If  the `bare coupling constant'
\m{g(\Lambda)} is  eliminated  in favor of \m{\mu}, all the
divergences will dissappear: this is the essence of renormalization. This
 kind of replacement of a (divergent) coupling constant by a momentum
scale is quite common in renormalization theory. It is sometimes called
`dimensional transmutation'.

Then \m{A_\Lambda} will have a limit as \m{\Lambda\to\infty}, and so
will the solution \m{\psi(\p)=\lim_{\Lambda\to \infty}\psi_\Lambda(\p)}:
\beq
	\psi(\p)={\chi(\p)\over {\p^2\over 2m}-E}+{1\over \phi(\mu,E)}
{1\over {\p^2\over 2m}-E}\int
{\chi(\q)\over {\q^2\over 2m}-E}[d\q].
\eeq
Here,
\beqs
	\phi(\mu,E)&=&\int[d\p]\bigg[{1\over {\p^2\over
2m}+{\mu^2\over 2m}}-{1\over {\p^2\over 2m}-E}\bigg]\cr
&=&2m\int{[d\p]\over \p^2}
\bigg[-{\mu^2\over \p^2+\mu^2}+{(-2mE)\over \p^2-2mE}\bigg].
\eeqs
This is a convergent integral. Note that the limiting solution is
independent of the choice of \m{\rho_\Lambda(\p)}.

Now, the resolvent kernel is given by the formula,
\beq
	\psi(\p)=\int R(E;\p,\q)\chi(\q)[d\p].
\eeq
Thus we get,
\beq
R(E;\p,\k)={(2\pi)^d\delta(\p-\k)\over {\p^2\over 2m}-E}+{1\over
\phi(\mu,E)}{1\over
{\p^2\over 2m}-E}{1\over {\k^2\over 2m}-E}.
\eeq

This is called the  `Krein formula' for the resolvent.  We can regard
our limiting system as defined by this formula for the
resolvent. The hamiltonian from which this follows is, {\em as a
differential operator,} the same as  that of the  free particle.
The interaction is
encoded into the boundary conditions at the  origin. The resolvent,
being a Green's function, encodes the information of these boundary
conditions as well. The function \m{{1\over \phi(\mu,E)}} gives thus a
convenient description of the interaction. We will  see that this is
(upto a constant) the  scattering amplitude.

Thus the interactions of the theory are described by the function 
\m{\phi(\mu,E)}.
Apart from it,  all the terms in the formula for the resolvent just
involve the free theory. We will see that with more than two
particles, we still have a similar formula, but the real valued
function \m{\phi(\mu,E)} is replaced by an operator valued
function,  the `principal operator'.  This principal operator acts on
a reduced  Hilbert space which for the simple case ofthis section is
one dimensional. That is why we just have a function \m{\phi(\mu,E)}
rather than an operator.

\subsection{ The Krein Formula and Boundary Conditions}

The resolvent of a differential operator 
contains the information  on its boundary conditions as well. It is
interesting to make explicit these boundary conditions implied by the
Krein formula for the hamiltonian. These boundary conditions determine
the domain of the hamiltonian thought of as a self-adjoint unbounded
operator in \m{L^2(R^3)}.

The domain of the hamiltonian is the range of its resolvent:
 the set of  functions  in momentum
space that can be written as  
\beq
	\int R(E;p,k)\chi(k)[dk]
\eeq
for square integrable  \m{\chi(k)}. Thus for
the free particle with resolvent
\beq
R_0(E;p,k)={(2\pi)^d\delta(p-k)\over {p^2\over 2m}-E}
\eeq
we have the set 
\beq
	\big\{\psi(p)={\chi(p)\over {p^2\over 2m}-E}|\chi\in L^2(R^3)\big\}. 
\eeq
In other words, the domain of the free hamiltonian consists of wave
 functions \m{\psi(p)} for which both  \m{\psi} and \m{p^2\psi} are  square
 integrable. In position space this means that the domain of the free
 hamiltonian is the set of wavefunctions that are square integrable
 and have square integrable second derivatives.

Now let us ask how the form of the resolvent changes if we change the
domain of the hamiltonian.  The difference of the resolvent from its value
 for a free particle is ( in
position space)
\beq
 \tilde R(E;x,y)-\tilde R_0(E;x,y)
\eeq
a homogenous solution of the Schr\"odinger equation: two different
Greens functions for the same differential equation differ by a
homogenous solution. The simplest possibility is to change the
behavior of the resolvent at one point ( say the origin) by putting
there  a pointlike 
scatterer. ( More complicated modifications are also allowed
mathematically, but are not as  interesting ). Then we must have, for \m{d=3},
\beq
\tilde R(E;x,y)=\tilde R_0(E;x,y)+C(E){e^{i\surd(2mE)|x|}\over |x|}
{e^{-i\surd(2mE)|y|}\over
|y|}
\eeq
for some constant \m{C(E)}. We require that the additional term corespond to
outgoing waves at spatial infinity, a physical requirement.  The formula for arbitrary \m{d} is similar
and involves Hankel functions. This means the wavefunction can blow up at
the origin as \m{C(E)\over |x|^{-1}}; (or \m{\log|x|} when \m{d=2}). Note that such a singularity is
still square integrable.

In momentum space this is of the form 
\beq
R(E;\p,\k)={(2\pi)^d\delta(\p-\k)\over {\p^2\over 2m}-E}+B(E){1\over
{\p^2\over 2m}-E}{1\over {\k^2\over 2m}-E}.
\eeq
The quantity \m{B(E)} is determined by the condition that this operator be
in fact a resolvent:
\beq
	{R(E;p,k)-R(E';p,k)\over E-E'}=\int R(E;p,k')R(E',k',k)[dk].
\eeq
After some calculations we get
\beq
	{B^{-1}(E)-B^{-1}(E')\over E-E'}= \int [dk]\big\{
{1\over {k^2\over 2m}-E}{1\over {k^2\over 2m}-E'}\big\}
\eeq
Or,
\beq
	B^{-1}(E)-B^{-1}(E')= \int [dk]\big\{
{1\over {k^2\over 2m}-E}-{1\over {k^2\over 2m}-E'}\big\}
\eeq
But this is precisely what we got from renormalization: our quantity 
\m{B(E)} is just the inverse of the  \m{\phi(E)} we had previously.

Thus we see that the  renormalized hamiltonian is just the free
hamiltonian with a modified boundary condition on the wavefunctions 
at the origin of position space.

\subsection{The Scattering Amplitude}

Consider the Schr\"dinger equation for a free particle:
\beq
	{1\over 2m}\nabla^2\tilde{\psi}(x)=E\tilde\psi(x).
\eeq
Suppose that \m{E>0}.
If we require that the wavefunction be continuos everywhere in space,
we have the  usual plane wave solution:
\beq
	Ce^{ik\cdot x}
\eeq
with \m{E={k^2\over 2m}}. In momentum space this  corresponds to a
wavefunction 
\beq
	(2\pi)^d\delta(p-q).
\eeq
But this is not the only solution if we allow the solution to blow up
at one point (say the origin). This would mean that there is a static
scatterer of zero size sitting at the origin. We would still require
the Schr\"odinger equation to hold away from the origin. Hence the
solution would have to differ from the plane wave by a multiple of a
homogenous solution of the differential equation, one that may diverge
at the origin. This homogenous solution represents the scattering by
the particle at the origin; hence we should require that it become an
outgoing wave at infinity:
\beq
	\tilde \psi(x)=e^{ik\cdot x} +f(k){e^{ikr}\over r}.
\eeq
Here the constant \m{f(k)} has the meaning of  the scattering amplitude.
In momentum space this becomes
\beq
\psi(\p)=(2\pi)^d\delta(\p-\k)+f(k){2\pi\over m}{1\over
{\p^2\over 2m}-{\k^2\over 2m}-i\eps}
\eeq
it being understood as usual that \m{\eps\to 0+}.

 We can now compare
this with what we get from the formula for the resolvent and see that
the extra term in the resolvent is precisely of this form:
\beq
\psi(\p)=(2\pi)^d\delta(\p-\k)+{1\over \phi(\mu,{\k^2\over 2m}+i\eps)}{1\over
{\p^2\over 2m}-{\k^2\over 2m}-i\eps}.
\eeq
This is in line with our argument that the Krein formula for the
resolvent describes a boundary condition on the wavefunction at the
origin: it is allowed to blow up.

Thus we have a scattering amplitude that is independent of angles:
\beq
	f(k)={m\over 2\pi \phi(\mu,{\k^2\over 2m}+i\eps)}.
\eeq
Evaluating the integral,
\beq
	\phi(\mu,E)={m\over 2\pi}\bigg[\surd(-2mE-i\eps)-\mu\bigg]
\eeq
so that the scattering amplitude is
\beq
	f(k)={1\over -\mu-ik}
\eeq
The  sign  is fixed by the rule \m{\lim_{\eps\to 0}(-1-i\eps)^{1/2}=-i}.

Thus we find exactly the `simple pole model' for the scattering amplitude for
low momenta that we
had in the last section, with scattering length \m{\xi={1\over \mu}}.

Later on we will study the many body problem of particles with such
contact interactions. We will give a complete theory only in
the simpler case of  two dimensional space. Problems with the
extension to the three dimensional case will be described as well.

\section{The Fermions with a Static Source}

Since our fermionic variant of the Lee model is simpler we will
describe it first.

Let \m{\psi(p),\psi^{\dag}(p)} be the creation-annihilation operators
for a fermion  field in {d+1}-dimensional space-time. They are  represented on
the  Fock space
\m{{\cal F}} built from the vacuum $|0>$:
\beq
[\psi(p),\psi^{\dag}(q)]_+=(2\pi)^d\delta(p-q),\quad
[\psi^{\dag}(p),\psi^{\dag}(q)]_+=0=[\psi(p), \psi(q)]_+\quad
\eeq
and
\beq
\psi(p)|0>=0.
\eeq
We are mainly interested in the case \m{d=3} but our method applies
 for any \m{d<4}.

On \m{{\cal F}\otimes C^2}, define the hamiltonian
\beq
	H_\Lambda=H_0+H_{1\Lambda}
\eeq
where,
\beq
	H_0=\int
[dp]\psi^{\dag}(p)\psi(p)\omega(p)
\eeq
and
\beq
	H_{1\Lambda}=\mu_\Lambda{1-\sigma_3\over 2}+g\int
[dp][\rho_\Lambda(p)\psi(p)\sigma_-+h.c.].
\eeq
There is no loss in assuming that the coupling constant \m{g>0} since any phase
in \m{g} can
be absorbed into a redefinition of the field \m{\psi(p)}. The quantity
\m{\mu_\Lambda} is a `bare coupling constant` whose dependence on
\m{\Lambda} will be determined later by renormalization.

The dispersion relation of the fermion field is chosen to  be non-relativistic:
\beq
	\omega_p=m+{p^2\over 2m}.
\eeq
The Pauli   matrices \m{\sigma_\pm,\sigma_3} act on \m{C^2} in the
usual way. There is a \m{U(1)} symmetry (rather like isopin) which
leads to the conserved quantity
\beq
	Q={1-\sigma_3\over 2}+\int[dp]\psi^{\dag}\psi(p).
\eeq
Clearly, \m{Q\geq 0}.

This model describes the interaction of the fermions with some heavy
particles sitting at the origin.
The function \m{\rho_{\Lambda}(p)} is a sort of form-factor:
it describes the  the internal structure of these heavy particles. For
us it
will serve the purpose of an ultra-violet regulator, eg. we may choose
\beq
	\rho_{\Lambda}(p)=\theta(|p|<\Lambda)
\eeq
for some momentum cut-off \m{\Lambda}. In the limit \m{\Lambda\to
\infty} we will have  point-like  heavy particles. But  in this limit there is
a  UV  divergence in the
theory. This divergence   can be removed by renormalizing  the constant
\m{\mu_{\Lambda}}. The final answer will not depend on the choice of
the form-factor \m{\rho_{\Lambda}}: any function which is equal to one at the
origin
and falls of faster than any power at infinity will give the same answers.
This is part of the universality of the transfinite theory: its
independence on the regularization scheme.

Our theory  can be thought of as a model for the interaction of the
top and charm  quarks with the Higgs boson: the heavy boson is the
Higgs boson, the heavy fermion the top quark and light fermion the
charm quark. Our model is an
approximation  where
\m{m_t,m_H>>m_c=m} but \m{|m_t-m_H|<<m_c}. In this limit we would have
the interaction  of  non--relativistic b-quarks with either a Higgs
boson or a top quark; the parameter \m{\mu} describes the mass
difference betwen the two heavy particles.

It is certainly true that
that \m{m_t>>m_c} and it is almost certain that \m{m_H>>m_c}. It
would be unusual for  the Higgs boson and the top  quark to have almost
the same mass, as we are asuming. Also, the Higgs-top-charm coupling
is very small in the standard model (induced by higher loop effects)
but is not so small in some variants of the standard model (for
example with supersymmetry). Thus our model describes a somewhat
unusual  possibility in particle physics:
but this has not yet been
ruled out experimentally.  In any case we can use this system as a physical
picture that guides our mathematical analysis of this renormalization
problem, much like the nucleon-pion system in the Lee model. More
realistic cases can be studied later, once the principles are
established in this simple case.

In some extensions of the standard model (eg. the minimal
supersymmetric  exension) there are charged Higgs particles. In these
models, a decay of the top quark into a Higgs boson and a bottom quark
is possible. Again, if the Higgs and the top quark happen to be almost
degenerate,  our model will describe this system.

The hamiltonian can be expressed as a \m{2\times 2} block split
according to \m{C^2}:
\beq
	H_\Lambda=\pmatrix{H_0&g\int [dp] \rho_{\Lambda}(p)\psi^{\dag}(p)\cr
                   g\int [dp] \rho_{\Lambda}(p)\psi(p)&H_0+\mu_{\Lambda}}.
\eeq
We will now construct the resolvent of this hamiltonian using some
identities given in the appendix. Then we will show that the limit as
\m{\Lambda\to\infty} of the resolvent exists: this is
renormalization. After that we will study
the spectrum of the renormalized theory.

In the appendix, we work out some formulae for inverting operators
split into \m{2\times 2} blocks as above. In the notation used there,
the resolvent iof the regularized hamiltonian s 
\beq
	R_\Lambda(E)={1\over H_{\Lambda}-E}==\pmatrix{\alpha&\beta^{\dag}\cr
		       \beta&\delta}
\eeq
where,
\beq
	\alpha={1\over H_0-E}+{1\over
H_0-E}b^{\dag}\Phi_\Lambda (E)^{-1}b{1\over H_0-E},\quad
\beta=-{1\over H_0-E}b\Phi_\Lambda (E)^{-1}
\eeq
and
\beq
	\delta=\Phi_\Lambda(E)^{-1},\quad
 b=g\int[dp]\rho_{\Lambda}(p)\psi(p).
\eeq
We express everything in terms of \m{\Phi_\Lambda(E)} since we will see that
all
the divergences are removed once we have a proper definition for
\m{\Phi_\Lambda(E)}. Once we know \m{\Phi}, the resolvent is given by
the above explicit formula.

Indeed, we have,
\beq
	\Phi_\Lambda(E)=H_0-E+\mu_{\Lambda}-g^2\int
[dp][dq]\rho_{\Lambda}(p)\rho_{\Lambda}(q)\psi(p){1\over H_0-E}\psi^{\dag}(q).
\eeq
Now we normal-order the  last term; i.e., we reorder the operators
so that the creation operators stand to the left of the annihilation
operators. This can be done using the identities
\beq
	H_0\psi^{\dag}(q)=\psi^{\dag}(q)[H_0+\omega(q)],\quad
\psi(p)\psi^{\dag}(q)=-\psi^{\dag}(q)\psi(p)+(2\pi)^d\delta(p-q),
\eeq
Thus
\beqs
	\Phi_\Lambda(E)&=&H_0-E+g^2\int
[dp][dq]\rho_{\Lambda}(p)\rho_{\Lambda}(q)\psi^{\dag}(p){1\over
H_0+\omega(p)+ \omega(q)-E}\psi(q)\cr
& &+\mu_{\Lambda}-g^2\int [dp]\rho_{\Lambda}^2(p){1\over H_0+\omega(p)-E}
\eeqs

Suppose we now choose
\beq
	\mu_\Lambda=\mu+g^2\int {\rho_\Lambda(p)^2[dp]\over \omega(p)-\mu}.
\eeq
Here \m{\mu} is a parameter independent of \m{\Lambda}. Moreover we
will require that \m{m>\mu} for simplicity. (This is not essential: it
guarantees that both the heavy states are stable against decay.)

The point of this choice of \m{\mu_\Lambda} is that it cancels the
divergent part of \m{\Phi_\Lambda(E)}:
\beqs
	\Phi_\Lambda(E)&=&H_0-E+g^2\int
[dpdq]\psi^{\dag}(p){\rho_\Lambda(p)\rho_\Lambda(q)\over
	            H_0+\omega(p)+\omega(q)-E}\psi(q)\cr
            & & -g^2\int [dp]\rho_\Lambda(p)^2\big[{1\over H_0+\omega(p)-E}-
{1\over \omega(p)-\mu}\big]+\mu
\eeqs

The  integrand of the last term behaves for large \m{|p|} like
 \m{\int { [dp]\over \omega(p)^2}}: it is
convergent if \m{d<4}, since  \m{\omega=m+{p^2\over 2m}}.
Thus we can take the limit as \m{\Lambda\to \infty} keeping \m{\mu}
fixed at a positive value less than \m{m}. Then, the limiting operator
\beqs
	\Phi(E)&=&H_0-E+g^2\int [dpdq]\psi^{\dag}(p){1\over
	            H_0+\omega(p)+\omega(q)-E}\psi(q)\cr
            & & -g^2\int [dp]\big[{1\over H_0+\omega(p)-E}-
{1\over \omega(p)-\mu}\big]+\mu
\eeqs
exists. This is the renormalized form of the principal operator.
W can just put this into the earlier formula to get a formula for the resolvent.

\subsection{The Principal Operator}

To get more explicit expressions we specialize to the case \m{d=3}. (In
fact the following  arguments apply to any \m{d<4}.) Evaluating
the integral,
\beqs
\Phi(E)&=&H_0-E+\mu+4\pi^2g^2(2m)^{3\over 2}\big[\surd
(H_0+m-E)-\surd(m-\mu)\big]\cr
    & &+g^2\int [dpdq]\psi^{\dag}(p){1\over H_0+\omega(p)+\omega(q)-E}\psi(q)
\eeqs
This principal  operator can be written as the sum of a `kinetic' (single
particle) term
\beq
	K(E)=H_0-E+\mu+4\pi^2g^2(2m)^{3\over 2}\big[\surd
(H_0+m-E)-\surd(m-\mu)\big]
\eeq
and an `interaction'  term
\beq
	U(E)=g^2\int [dpdq]\psi^{\dag}(p){1\over
H_0+\omega(p)+\omega(q)-E}\psi(q).
\eeq
The fermion number operator \m{N=\int
[dp]\psi^{\dag}(p)\psi(p)} commutes with \m{\Phi(E)}.

The principal operator determines the dynamics of the theory. The
resolvent of the hamiltonian of the renormalized theory is
\beq
	R(E)=\pmatrix{\alpha(E)&\beta^{\dag}(E)\cr
		       \beta(E)&\Phi(E)^{-1}}
\eeq
where
\beq
	\alpha(E)={1\over H_0-E}+{g\over H_0-E}\int
[dq]\psi^{\dag}(q){1\over \Phi(E)}\int [dp]\psi(p){g\over H_0-E}
\eeq
and
\beq
	\beta(E)=-{g\over \Phi(E)}\int [dp]\phi(p){1\over H_0-E}.
\eeq
This can be thought of as a sort of Krein formula for the resolvent of
our field theory.

Indirectly it defines the hamiltonian \m{H} as the
operator for which \m{R(E)={1\over H-E}} is the resolvent. But if we
were to think of this renormalized hamiltonian directly as an operator on the
fermionic Fock space, it would appear   to be just the free
fermion operator.  For example,  consider  the space of states
\beq
	|f>=\int f(p_1\cdots p_n)\psi^{\dag}(p_1)\cdots \psi^{\dag}(p_n)
[dp_1\cdots dp_n]|0>
\eeq
with  smooth \m{f(p_1,\cdots p_n)} decreasing at infinity faster than  any
polynomial. They correspond to fermionic wavefunctions that are smooth
in position space that also fall of faster than any polynomial.
 On these `nice' states,
the hamiltonian \m{H} of the transfinite theory is the
same
the free hamiltonian:
\beq
	H|f>=\int [\omega(p_1)+\cdots \omega(p_n)]f(p_1\cdots
p_n)\psi^{\dag}(p_1)\cdots \psi^{\dag}(p_n)|0>
[dp_1\cdots dp_n].
\eeq

But this  formula does not uniquely define \m{H} as a self-adjoint operator. It
has many self-adjoint extensions, corresponding to different sets of
boundary conditions as the momenta go to infinity. One of them is the
free hamiltonian \m{H_0}, but there are other extensions too which
define  interacting theories.
The interactions are hidden in the domain of
definition of this unbounded operator: in other words in the boundary
conditions of the fermion wavefunctions in the limit of large
momentum.
 It is quite awkward to think
this way: for example  it would be difficult to calculate the
spectrum of the renormalized theory.

But the same information is
contained in a more explicit form in the above Krein formula for the
resolvent. The resolvent is a sort of Green's function which therefore
contains also the information on the boundary conditions. Another way
to understand it is that (away from the spectrum) the resolvent is a
bounded operator, so there is no need to specify its domain.

Transfinite field theories are thus theories in which the interaction
takes place at infinity  in momentum space. We are still able to give
a sensible description of these theories.

\subsection{The  Spectrum of the Transfinite Theory}

We saw that there is an explicit formula for the resolvent of the
hamiltonian in terms of the inverse of the Principal operator.
Thus, all the dynamical information about the theory is contained in its
principal operator. For example the discrete spectrum  of the hamiltonian
correspond to the poles of the resolvent. There are no poles in
\m{1\over H_0-E}: its spectrum is purely continuous. Thus the poles  must arise from those of
\m{\Phi(E)^{-1}}; i.e., roots of the equation
\beq
	\Phi(E)|u>=0
\eeq
This equation  now plays the role of the Schrodinger eigenvalue problem in the
transfinite theory.
The residue of the pole of the resolvent is the projection operator to
the corresponding eignspace of \m{H}. Thus the eigenvector of \m{H}
corresponding to a  root \m{E_0} of the principal operator   is given
by
\footnote{This root is assumed not be in the spectrum of \m{H_0}. In
this case it would be be embedded in  the continuous spectrum of \m{H}
and should be unstable. We will deal with the continuous spectrum of
\m{H} later.}
\beq
 \pmatrix{{g\over H_0-E_0}\int [dp]\psi^{\dag}(p)|u>\cr |u>}.
\eeq
This is because the residue of a pole in the resolvent is the
projection operator to the eigenspace with that eigenvalue. We can
read off  this residue and see that it is the projection to
the above state, once a root of the equation \m{\Phi(E)|u>=0} is found.

An example is the vacuum state:
\beq
	\Phi(E)|0>=\big(-E+\mu+g^2(2m)^{3\over 2}[\surd(m-E)-\surd(m-\mu)]\big)|0>.
\eeq
There  is a root when
\beq
	E=\mu.
\eeq
Since \m{\mu<m}, it is not in  the spectrum of \m{H_0}.
The corresponding eigenvector of the hamiltonian is
\beq
	\pmatrix{{g\over H_0-\mu}\int [dp]\psi^{\dag}(p)|0>\cr
			|0>}.
\eeq

 It contains a fermion
in the first entry, so it is not the vacuum in the whole Hilbert space
\m{{\cal F}\otimes  C^2}. In spite of this, it is the state of lowest
energy-the ground state- when \m{\mu<0}. (Proof is in the next
section). When \m{m>\mu>0} the above state is the first excited state.( The
ground state is  just the `vacuum' state \m{\pmatrix{|0>\cr 0}}.)
When \m{\mu>m}, we don't get an eigenstate of the hamiltonian since
this state is then unstable. (The ground state in this case is
also\m{\pmatrix{|0>\cr 0}}.)

 More generally, the spectrum of the hamiltonian is the set
of values of \m{E} at which the resolvent either does not exist
(discrete spectrum) or exists but is unbounded (continuous
spectrum). Thus the continuous spectrum will be that of \m{H_0} plus
 the values of \m{E} at which \m{\Phi(E)} does not have a
bounded inverse.

\subsection{Proof That The Ground State Energy is Finite}

In order to see that we have exorcised all the infinities, we must
show not only  that the resolvent of the hamiltonian
exists in the limit as \m{\Lambda\to
\infty}  but also that the ground state energy is finite.
This is nontrivial to prove since we will see that there are many
theories where even after a renormalization there are further
divergences which make the spectrum not bounded below.

We will estimate the norm \m{||\Phi(E)^{-1}||} for the case \m{E<\mu};
our aim is to show that this is finite. Then the ground state energy is
either zero (when \m{\mu>0})  or is equal  to \m{\mu}.

 It
is sufficient to consider the sector with the fermion number  held
fixed at some value \m{n}, since \m{\Phi(E)} preserves this number.

Recall that \m{\Phi(E)=K(E)+U(E)} with
\beq
	K(E)=H_0-E+\mu+4\pi^2g^2(2m)^{3\over 2}\big[\surd
(H_0+m-E)-\surd(m-\mu)\big]
\eeq
and
\beq
	U(E)=g^2\int [dpdq]\psi^{\dag}(p){1\over
H_0+\omega(p)+\omega(q)-E}\psi(q).
\eeq
Moreover,
\beq
	K(E)\geq nm+(\mu-E),\quad U(E)\geq 0.
\eeq
The inequalities become equalities when the fermion number is zero.

Now write,
\beq
	\Phi(E)=K(E)^{\half}\big[1+\tilde U(E)\big]K(E)^{\half},\quad
 \tilde U(E)=K(E)^{-\half}U(E)K(E)^{-\half}.
\eeq
Since \m{\tilde U(E)\geq 0},
\beqs
	||\Phi(E)^{-1}||&\leq& ||K(E)^{-1}||\;\;||\big[1+\tilde
U(E)\big]^{-1}||\cr
& \leq&
\big[nm+(\mu-E)\big]^{-1}\cr
&\leq& {1\over \mu-E}.
\eeqs
This shows that \m{\Phi(E)} has spectrum bounded below  by \m{\mu}.

This simple method is not sufficient to determine the ground state 
 in the bosonic case. We will have to supplement it
with a  sort of mean field theory.

\section{The Lee Model}

The   Lee  model  has a charged
bosonic field satisfying:
\beq
[\phi(p),\phi^{\dag}(q)]=(2\pi)^d\delta(p-q),\quad
[\phi^{\dag}(p),\phi^{\dag}(q)]=0=[\phi(p), \phi(q)].
\eeq
The bosonic Fock space \m{{\cal B}} is built from the vacuum in the
usual way:
\beq
\phi(p)|0>=0.
\eeq
Again, the complete Hilbert space of the system is \m{{\cal B}\otimes
C^2}.

The  hamiltonian is, again,
\beq
H_{\Lambda}=H_0+H_{1\Lambda}
\eeq
with
\beq
	H_0=\int
[dp]\phi^{\dag}(p)\phi(p)\omega_p
\eeq
and
\beq
	H_{1\Lambda}=\mu_\Lambda{1-\sigma_3\over 2}+
g\int [dp][\rho_\Lambda(p)\phi(p)\sigma_-+h.c.].
\eeq
Some of the methods are the same as in the fermionic case; we will
then omit the details. But the bosonic case requires some new analysis
as well.

The divergences are removed as before by normal ordering and choosing the bare
mass difference to be
\beq
	\mu_\Lambda=\mu+g^2\int {\rho_\Lambda(p)^2[dp]\over \omega(p)-\mu}.
\eeq
In the limit the resolvent has the form:
\beq
 R(E)=\lim_{\Lambda\to\infty}{1\over
H_{\Lambda}-E}=\pmatrix{\alpha(E)&\beta^{\dag}(E)\cr
		       \beta(E)&\Phi(E)^{-1}};
\eeq
\beq
	\alpha(E)={1\over H_0-E}+{1\over H_0-E}\int
[dq]\phi^{\dag}(q){1\over \Phi(E)}\int [dp]\phi(p){1\over H_0-E};
\eeq
\beq
	\beta(E)=-{1\over \Phi(E)}\int [dp]\phi(p){1\over H_0-E};
\eeq
and finally
\beqs
\Phi(E)&=&H_0-E+\mu+4\pi^2g^2(2m)^{3\over 2}\big[\surd
(H_0+m-E)-\surd(m-\mu)\big]\cr
    & &-g^2\int [dpdq]\phi^{\dag}(p){1\over H_0+\omega(p)+\omega(q)-E}\phi(q).
\eeqs
Other than the sign of the last term the answer is essentially the
same as before. But this sign makes an important difference.

The proton is the state
\m{\pmatrix{|0>\cr 0}} and it is an eigenstate of the renormalized
hamiltonian with eigenvalue zero. The neutron
is an  eigenstate of the hamiltonian,
\beq
	\pmatrix{{g\over H_0-\mu}\int [dp]\phi^{\dag}(p)|0>\cr
			|0>}.
\eeq
 It corresponds the root of
\beq
	\Phi(E)|0>=\big(-E+\mu+g^2(2m)^{3\over 2}[\surd(m-E)-\surd(m-\mu)]\big)|0>
\eeq
with  \m{E=\mu}. Thus the parameter \m{\mu} is the renomalized value
of the neutron-proton mass difference. So it is reasonable to assume
\m{m>\mu>0}.

But is the proton the state of least energy? Could there be states which
contain many bosons that have a lower energy?. Is there even a ground
state?
 These questions can be answered by studying the
principal operator \m{\Phi(E)}.

\subsection{ Lower Bound for the Ground State Energy }

We will estimate the norm \m{||\Phi(E)^{-1}||} for the case \m{E<\mu}. It
is sufficient to consider the sector with the number of bosons held
fixed at \m{n}, since \m{\Phi(E)} preserves this number.

Let us define  the ``kinetic'' part  of \m{\Phi(E)} to be:
\beq
K(E)= H_0-E+\mu+4\pi^2g^2(2m)^{3\over 2}\big[\surd (H_0+m-E)-\surd(m-\mu)\big].
\eeq
Clearly,
 \beq
	K(E)\geq nm+(\mu-E).
\eeq

Define \m{-U(E)} to be the ``potential'' part of \m{\Phi(E)}:
\beq
	\Phi(E)\equiv K(E)-U(E).
\eeq
Notice that \m{\Phi(E)} is now the difference of two positive oeprators
rather than the sum as in the fermionic case. This is what makes the
bosonic case more complicated.

As before,
\beq
	\Phi(E)=K(E)^{\half}\big[1-\tilde U(E)\big]K(E)^{\half}
\eeq
where
\beq
	\tilde U(E)=K(E)^{-\half}U(E)K(E)^{-\half}.
\eeq
Also,
\beq
	||\Phi(E)^{-1}||\leq ||K(E)^{-1}||\;\;||\big[1-\tilde U(E)\big]^{-1}||;
\eeq
i.e.,
\beq
	||\Phi(E)^{-1}||\leq \big[nm+(\mu-E)\big]^{-1}||\big[1-\tilde
U(E)\big]^{-1}||.
\eeq
Now, by explicit calculation,
\beq
\tilde U(E)=g^2\int [dpdq]\phi^{\dag}(p)
\bigg[{1\over
K(E-\omega_p)^\half\big(H_0+\omega_p+\omega_q-E\big)
K(E-\omega_q)^\half}\bigg]\phi(q).
\eeq
Thus, (remembering that inside the square bracket the boson number is \m{n-1}),
\beqs
\tilde U(E)&\leq& g^2\int [dpdq]\phi^{\dag}(p)
\big[(n-1)m+\mu-E+\omega_p\big]^{-\half}\cr
& & \big[(n-1)m+\omega_p+\omega_q-E\big]^{-1}\cr
& &
\big[(n-1)m+\mu-E+\omega_q\big]^{-\half}\phi(q).
\eeqs
 Also, in the sector with
\m{n} bosons,
\beq
||\int \phi^\dag(p)u(p,q)\phi(q)[dpdq]||\leq n\bigg[\int
|u(p,q)|^2[dpdq]\bigg]^{\half}.
\eeq
Combining these,
we get
\beqs
	||\tilde U(E)||^2&\leq& g^4n^2
\int [dpdq]\big[(n-1)m+\mu-E+\omega_p\big]^{-1}\cr
& & \big[(n-1)m+\omega_p+\omega_q-E\big]^{-2}\cr
& &
\big[(n-1)m+\mu-E+\omega_q\big]^{-1} .
\eeqs
Now we put in \m{\omega(p)={p^2\over 2m}+m}.In  the middle factor  we
can replace an \m{m} by a \m{\mu} since \m{m>\mu}:
\beq
	\big[(n-1)m+2m+{p^2\over 2m}+{q^2\over 2m}-E\big]^{-2}<\big[nm+\mu+{p^2\over
2m}+{q^2\over 2m}-E\big]^{-2}.
\eeq
This makes all the constants in the denominators the same so that we
can scale them out to get
\beqs
	||\tilde U(E)||^2&\leq& g^4n^2{(2m)^3\over [nm+\mu-E]}\cr
& & \int {[dpdq]\over (1+p^2)(1+q^2)(1+p^2+q^2)^2} .
\eeqs
The integral is convergent.
Working our way back to the beginning,
\beq
	||\Phi(E)||^{-1}\leq {1\over \surd(nm+\mu-E)}
         {1\over \bigg[\surd(nm+\mu-E)-g^2n(2m)^{3\over 2}C\bigg]}
\eeq
where
 \beq
C^2=\int {[dpdq]\over (1+p^2)(1+q^2)(1+p^2+q^2)^2 }.
\eeq

If \m{E} is to be an eigenvalue of energy, it must be big enough to
 make the denominators on the r.h.s. vanish; otherwise,
 \m{\Phi(E)^{-1}} would remain bounded. This means there is a lower
 bound on all eigenvalues, which really is the same as a lower bound
 on the ground state energy:
\beq
	E_{gr}\geq (nm+\mu)-n^2g^4(2m)^3C^2.
\eeq
 Thus we see that in each sector with a fixed
number of bosons, there is a ground state. However, there is still the
possibility that the ground state energy diverges as \m{n} grows to
infinity. In this limit we should expect all the bosons to settle into
the same state. The ground state of the free theory has all the bosons
 in the zero momentum state. In the interacting theory, the ground
 state could in general be something quite different. For example, the
 bosons might settle into a state which is concentrated at the
 origin. The difference between that bosonic number density and its
 free field value is called the `boson condensate'. 
Whether such a non--zero condensate  of
bosons  forms   cannot be settled by the
present analysis: we need to study the limit as \m{n\to \infty}.

\subsection{The Large \m{n} Limit of the Lee Model}

In the limit that the number of bosons becomes large, we should be
able to use   mean field theory. We would expect
all the bosons to occupy the same state \m{u(p)}.
This state is normalized so that the occupation number is \m{n}:
\beq
	||u||^2=\int |u(p)|^2[dp]=n.
\eeq
In the  limit \m{n\to\infty}, operators can be approximated by their
expectation values in this state: a kind of mean field theory. They
will then become functions on the space of such states, the complex
projective space\footnote{There is no physical effect if \m{u} is
replaced by \m{e^{i\theta}u}. The space of vectors of fixed length,
modulo  a phase, is complex projective space.} of \m{L^2(R^3)}.

Thus our principal operator becomes the {\em principal function}
\beqs
\Phi(E,u)&=&h_0(u)-E+\mu+4\pi^2g^2(2m)^{3\over 2}\big[\surd
(h_0(u)+m-E)-\surd(m-\mu)\big]\cr
    & &-g^2\int [dpdq]{u^*(p)u(q)\over h_0(u)+\omega(p)+\omega(q)-E}.
\eeqs
Here,
\beq
	h_0(u)=\int\omega(p)|u(p)|^2[dp].
\eeq
We must solve \m{\Phi(E,u)=0} to get \m{E} as a function of
\m{E}. Then we must find the \m{u} that gives the smallest such \m{E},
subject to the constraint on the norm \m{||u||^2=n}.

It is convenient  to reexpress the problem in terms of some new
variables. The  principal function depends on \m{E} only through the
combination
\beq
	\lambda=\int \omega(p)|u(p)|^2 [dp]-E.
\eeq
So define,
\beqs
f(\lambda,u)&=&\lambda+\mu+(2\pi g)^2(2m)^{3\over
2}[\surd(\lambda+m)-\surd(m-\mu)]\cr
& & -g^2\int [dpdq]{u^*(p)u(q)\over
\lambda+\omega(p)+\omega(q)}.
\eeqs
Then \m{\lambda} is determined by the equation \m{f(\lambda,u)=0} and
\m{E} by
\beq
	E=nm+{1\over 2m}\int p^2|u(p)|^2[dp]-\lambda.
\eeq

Putting in the explicit form of \m{\omega(p)},
\beqs
f(\lambda,u)&=&\lambda+\mu+(2\pi g)^2(2m)^{3\over
2}[\surd(\lambda+m)-\surd(m-\mu)]\cr
& & -2mg^2\int [dpdq]{u^*(p)u(q)\over
  p^2+q^2+2m(\lambda+2m)}.
\eeqs
We now make the change of variables
\beq
u(p)=\surd n\ \  [2m(2m+\lambda)]^{3\over 2}  v(\surd[2m(2m+\lambda)]p).
\eeq
The change of scale of momentum will simplify the denominator in the
integral for \m{f(\lambda,u)}. The overall factor of \m{\surd n} turns
the normalization condition into
\beq
	\int |v(p)|^2[dp]=1
\eeq
so that we can separate out the \m{n} dependence.

Now we have
\beq
	E=nm-\lambda+n(\lambda+2m)\int p^2|v(p)|^2[dp],
\eeq
and
\beqs
f(\lambda,u)&=&\lambda+\mu+(2\pi g)^2(2m)^{3\over
2}[\surd(\lambda+m)-\surd(m-\mu)]\cr
& & -ng^2(2m)^{3\over 2}\surd[\lambda+2m]\int [dpdq]{v^*(p)v(q)\over
  p^2+q^2+1}.
\eeqs
Imposing \m{f(\lambda,u)=0} will give an equation for \m{\lambda}:
\beqs
& & [\lambda+2m]^{-\half}\big\{\lambda+\mu+(2\pi g)^2(2m)^{3\over
2}[\surd(\lambda+m)-\surd(m-\mu)] \big\} =\cr
 & & ng^2(2m)^{3\over 2}\int [dpdq]{v^*(p)v(q)\over
  p^2+q^2+1}.
\eeqs
Then \m{\lambda} determined as a function of \m{v}. 
The l.h.s. can be seen to be a monotonically increasing  function of \m{\lambda}
 by writing it as
\beqs
\surd(\lambda+2m)&-&{m\over \surd(\lambda+2m)}+\cr
(2\pi g)^2(2m)^{3\over 2}\surd\big[1-{m\over
\lambda+2m}\big]&+&{\mu-m-(2\pi g)^2(2m)^{3\over 2}\surd(m-\mu)\over
\surd(\lambda+2m)}.
\eeqs
(Recall that \m{\mu<m}).
 Thus we can solve for \m{\lambda} in terms of the r.h.s. as a
monotonically increasing  function as well. Its minimum values is
\m{-\mu}, attanined when the r.h.s. is zero. Thus, we have 
\beq
	\lambda=-\mu+f_1(nU)
\eeq
where
\beq
	U=g^2(2m)^{3\over 2}\int [dpdq]{v^*(p)v(q)\over p^2+q^2+1}
\eeq
and \m{f_1} is   monotonically increasing. (We won't need an explicit
form for \m{f_1}.)
Now put this  it into
the expression for energy: 
\beq
	E=nm+\mu+(2m-\mu)K+f_1(nU)[nK-1]
\eeq
where
\beq
	K=\int [dp]p^2|v(p)|^2.
\eeq

Now suppose we replace \m{v(p)} by \m{v_a(p)=a^{-{3\over
2}}v(a^{-1}p)}. Let \m{K(v_a)=K_a(v),E(v_a)=E_a(v)} etc. 
Now, \m{K_a=a^{2}K} and \m{U_a=a^{4}U}: as a function
of \m{a} both \m{K_a} and \m{U_a} are increasing. 
Then \m{ E_a} is also a monotonically 
increasing function of \m{a}; the minimum for \m{E_a} will occur  at
\m{a=0} and \m{K,U= 0}. Thus we find that the  groundstate
energy is just
\beq
	E=nm+\mu.
\eeq
All the bosons are in the zero momentum state in the large \m{n}
limit: there is no condensate. In other words the ground state of the
bosonic field is essentially the same as in the free theory: the
interaction is not strong enough to modify the ground state
substantially.
We will see examples later where the ground state is affected by the 
interaction.

\section{The \m{\lambda\phi^4_{2+1NR}} Model }

So far we considered theories where the coupling constants did not
need to be renormalized. In fact the UV divergence  was removed by a
normal ordering of the principal operator. Now we will study a system
where the coupling constant needs to be renormalized: in fact one that
is asymptotically
free\cite{thorn,huang}\cite{jackiw}\cite{guptarajeev}
 We will discover a trick of introducing
fictitious degrees of freedom (`angels') which will help us reduce the
renormalization again to a normal ordering of the principal operator.

The theory of interest is the  nonrelativistic scalar field theory in
two space dimensions,
with   a \m{\lambda\phi^4} interaction. This is the many body problem
of non-relativistic bosons  interacting through a `delta function'
potential. It is well-known that, the two body problem
 has an ultra-violet divergence which can be removed by a coupling constant
renormalization.  In Ref. \cite{hendersonrajeev}  the `few body
problem'  was studied by quantum mechanical renormalization  methods.
In this section we will  renormalize this model  by viewing it as a
non-relativistic quantum field theory: in other words we will study
the many  body problem.  We will carry out the analysis in part  for
arbitrary \m{d}; this will show  why our renormalization method is
not sufficient for \m{d=3}.

Define, on the Bosonic Fock space \m{{\cal B}},
the regularized hamiltonian operator
\beq
	H_\Lambda=H_0+H_{1\Lambda}
\eeq
where
\beq
H_0=\int {p^2\over 2}\phi^{\dag}(p)\phi(p)[dp]
\eeq
and
\beqs
	H_{1\Lambda}&=&-g(\Lambda)\int[dp_1dp_2dp_1'dp_2']
\rho_{\Lambda}(p_1-p_2)\rho_\Lambda(p_1'-p_2')\cr
	& & 		(2\pi)^d\delta(p_1+p_2-p_1'-p_2')
     \phi^{\dag}(p_1)\phi^{\dag}(p_2)\phi(p_1')\phi(p_2').
\eeqs
Here
\beq
	\rho_\Lambda(p)=\theta(|p|<\Lambda)
\eeq
as before. The dimensionless constant \m{g(\Lambda)} is positive which
corresponds to an attractive interaction between the bosons. We have
shown elsewhere that \cite{hendersonrajeev} there is an ultra-violet divergence
 as \m{\Lambda\to \infty}, which can be removed by renormalizing the coupling
constant.
Our aim is         to get a manifestly
finite  expression for the resolvent of the renormalized quantum field
theory.

Now we will introduce a little trick that simplifies our problem: we 
will introduce new particles called angels that describe a bound state
of a pair of bosons. However, they are created by operators with
unusual defining relations: these are necessary  to avoid
over-counting the number of degrees of freedom.

\subsection{Angels}

Define operators  satisfying
\beq
	\chi(\p)\chi^{\dag}(\q)=(2\pi)^d\delta(\p-\q),\quad
\chi(\p)\chi(\q)=0=\chi^{\dag}(\p)\chi^{\dag}(\q).
\eeq
Note that it is the {\it product} and not the {\it commutator} that appears
here. These operators can be represented on the Hilbert space
\m{C\oplus L^2(R^d)}. We can regard \m{\chi^{\dag}(\p)} as creating an
entity (we will give it the somewhat whimsical name `angel') out of
the empty state represented by \m{C}. There can be at most one angel
in any state: that is the meaning of the product of the creation
operators being zero. (This is an extreme example of the exclusion
statistics considered in some other contexts.)

Now consider an augmentation of the Bosonic Hilbert space,\m{\tilde{\cal
B}={\cal
B}\oplus {\cal B}\otimes L^2(R^d)}.
On it define the hamiltonian
\beqs
	{\tilde H}_\Lambda&=&H_0\Pi_0+
 \int
[dp_1 dp_2 dp_3]\rho_\Lambda(p_1-p_2)\cr
& & \phi^{\dag}(p_1)\phi^{\dag}(p_2)\chi(p_3 )(2\pi)^d\delta(p_1+p_2-p_3)+h.c.\bigg]+{1\over g(\Lambda)}\Pi_1.
\eeqs
Here \m{\Pi_0} is the projection operator to the subspace containing
no angel :
\beq
	\Pi_0=\int [d\p]\chi(\p)\chi^{\dag}(\p),
\eeq
and \m{\Pi_1} the projection operator to the subspace with exactly one
angel:
\beq
	\Pi_1=\int [d\p]\chi^{\dag}(\p)\chi(\p).
\eeq

The point of introducing angels is this: the projection of the resolvent
of the \m{\tilde{H_\Lambda}},  to  the \m{\cal B} (the sector with no angels)
is
just the resolvent of the original bosonic system. But we will get
another formula for this resolvent, using angels,  which has a finite limit as
\m{\Lambda\to \infty}. Indeed we
will see that  the infinity is avoided by keeping the energy
of the bound state of a pair of bosons fixed in this limit;
 an angel is essentially   such a bound state.

Let us split the  Hilbert space according to the angel number, with a
corresponding splitting of the operator:
\beq
	{\tilde H}_\Lambda-E\Pi_0=\pmatrix{a&b^{\dag}\cr b&d\cr}
\eeq
with
\beq
	a:{\cal B}\to {\cal B}, \quad b^{\dag}:{\cal B}\otimes
L^2(R^d)\to {\cal B}, d:{\cal B}\otimes L^2(R^d)\to {\cal B}\otimes
L^2(R^d).
\eeq
Define  an operator \m{{\tilde R}_{\Lambda}(E)} split in the same
way:
\beq
	{\tilde R}_{\Lambda}(E)=
{1\over {\tilde H}_\Lambda-E \Pi_0}=\pmatrix{\alpha&\beta^{\dag}\cr
	      \beta&\delta}.
\eeq
We claim that
\beq
	\alpha={1\over H_\Lambda-E};
\eeq
i.e., {\em  \m{{\tilde R}_{\Lambda}} projected to
\m{{\cal B}} is just the resolvent of \m{H_\Lambda}}. To see this, we
use the formula we obtained earlier:
\beq
	\alpha=[a-b^{\dag}d^{-1}b]^{-1}.
\eeq
For us now \footnote{ We hope there will be no confusion in using the
same symbol \m{d} both for an operator and the dimension.},
\beqs
	a&=&H_0-E,\quad d={1 \over g(\Lambda)},\cr
 b^{\dag}&=&
\int
[dp_1dp_2]\rho_\Lambda(p_1-p_2)\phi^{\dag}(p_1)\phi^{\dag}(p_2)
\chi(p_1+p_2)
\eeqs
This gives
\beqs
	b^{\dag}d^{-1} b&=&{g(\Lambda)}
\int [dp_1 dp_2dp_3]\rho_\Lambda(p_1-p_2)\phi^{\dag}(p_1)
				\phi^{\dag}(p_2)\chi(p_3)\cr
& & (2\pi)^d\delta(p_1+p_2-p_3)\cr
& & 
\int [dp_1' dp_2' dp_3']\rho_\Lambda(p_1'-p_2')\chi^{\dag}(p_3')
\phi(p_1')\phi(p_2')
(2\pi)^d\delta(p_1'+p_2'-p_3')\cr
\eeqs
If we use
\beq
	\chi(\p)\chi^{\dag}(\p')=(2\pi)^d\delta(\p-\p')
\eeq
we will get the required result.

But we have another formula for this resolvent:
\beq
\alpha=a^{-1}+a^{-1}b^{\dag}[d-ba^{-1}b^{\dag}]^{-1}a^{-1}.
\eeq
This will give,
\beq
	{1\over H_\Lambda-E}=a^{-1}+ {1\over
           2}a^{-1}b^{\dag}\Phi_\Lambda(E)^{-1} b a^{-1}
\eeq
where
\beqs
\Phi_\Lambda(E)&=&{1\over g(\Lambda)}-
\int
[dp_1dp_2dp_1'dp_2']\rho_\Lambda(p_1-p_2)\rho_\Lambda(p_1'-p_2')\cr
& & 
\chi^{\dag}(p_1+p_2)\bigg[
  \phi(p_1)\phi(p_2){1\over H_0-E}
  \phi^{\dag}(p_1')\phi^{\dag}(p_2')\bigg]\chi(p_1'+p_2').
\eeqs
We can regard \m{b} as the operator that converts a pair of bosons
into an angel. Then \m{\Phi_\Lambda(E)} is a  kind of effective
hamiltonian in the sector with one angel and two fewer bosons: its
zeros are energy levels of the manybody problem.

Note that in this way
of writing the resolvent of \m{H_\Lambda}, the coupling constant
appears additively! Its renormalization can be done by separating out
a divergent constant from \m{\Phi_\Lambda(E)}. We will do this by normal
ordering the operators in \m{\Phi_{\Lambda}(E)}.

Using the canonical commutation relations and
\beq
	H_0\phi^{\dag}(\p)=\phi^{\dag}(\p)H_0+{\p^2\over 2}
\eeq
we can rewrite the quantity in the square brackets above equation  as
\beqs
\phi^{\dag}(p_1')\phi^{\dag}(p_2')
{1\over H_0+\omega(p_1')+\omega(p_2')+\omega(p_1)+\omega(p_2)-E}
		\phi(p_1)\phi(p_2)\cr
+(2\pi)^d\delta(p_1-p_1')\phi^{\dag}(p_2'){1\over
H_0+\omega(p_1')+\omega(p_2')+\omega(p_2)-E}\phi(p_2)\cr
+(2\pi)^d\delta(p_1-p_2')\phi^{\dag}(p_1'){1\over
H_0+\omega(p_1')+\omega(p_2')+\omega(p_2)-E}\phi(p_2)\cr
+(2\pi)^d\delta(p_2-p_1')\phi^{\dag}(p_2'){1\over
H_0+\omega(p_1')+\omega(p_2')+\omega(p_1)-E}\phi(p_1)\cr
+(2\pi)^d\delta(p_2-p_2')\phi^{\dag}(p_1'){1\over
H_0+\omega(p_1')+\omega(p_2')+\omega(p_1)-E}\phi(p_1)\cr
(2\pi)^d\delta(p_1-p_1')(2\pi)^d\delta(p_2-p_2')
{1\over H_0+\omega(p_1')+\omega(p_2')-E}\cr
(2\pi)^d\delta(p_1-p_2')(2\pi)^d\delta(p_2-p_1')
{1\over H_0+\omega(p_1')+\omega(p_2')-E}\cr
\eeqs
This  gives, 
\beqs
   \Phi_\Lambda(E)&=&g^{-1}(\Lambda)-
\int
[dp_1dp_2dp_1'dp_2']\rho_\Lambda(p_1-p_2)\rho_\Lambda(p_1'-p_2')\cr
& & 
\chi^{\dag}(p_1+p_2)\bigg[\cr
& & \phi^{\dag}(p_1')\phi^{\dag}(p_2')
{1\over H_0+\omega(p_1')+\omega(p_2')+\omega(p_1)+\omega(p_2)-E}
		\phi(p_1)\phi(p_2)\cr
& & +4(2\pi)^d\delta(p_1-p_1')\phi^{\dag}(p_2'){1\over
H_0+\omega(p_1')+\omega(p_2')+\omega(p_2)-E}\phi(p_2)\cr
& & +2(2\pi)^d\delta(p_1-p_1')(2\pi)^d\delta(p_2-p_2')
{1\over H_0+\omega(p_1')+\omega(p_2')-E}\bigg]\cr
& & \chi(p_1'+p_2').
\eeqs

Now we can take the limit as \m{\Lambda\to \infty}. The only divergent
 term is the last one. If we choose for \m{g(\Lambda)}
the expression from the two-body problem,
\beq
	g^{-1}(\Lambda)={4\over 2^d}\int \rho^2_\Lambda(\p){1\over \p^2+\mu^2}
\eeq
this divergence will cancel yielding
 a finite expression for \m{\Phi_\Lambda(E)} as
\m{\Lambda\to \infty}.
In fact, we have,
\beq
\lim_{\Lambda\to\infty}\int [dq]
\rho^2_\Lambda(\q)[ d\q]
\bigg[{1\over {\q^2\over 2}+{\mu^2\over 2}}-{1\over
H_0+{\p^2+\q^2\over 2}-E}\bigg]= \xi({\mu^2\over 2},H_0+{p^2\over 2}-E).
\eeq
Here,
\beqs
	\xi({\mu^2\over 2},{\nu^2\over 2})&:=&{4\over 2^d}\int[ d\p]
\bigg[{1\over p^2+\mu^2}-{1\over p^2+\nu^2}\bigg]\cr
&=& {4\over 2^d}\bigg[\int[ d\p]
{\nu^2\over p^2(p^2+\nu^2)}-\int[d\p]{\mu^2\over p^2(p^2+\mu^2)}\bigg]\cr
&:=&\xi({\nu^2\over 2})-\xi({\mu^2\over 2})
\eeqs

We will get
\beqs
   \Phi(E)&=&\int [d\p]\chi^{\dag}(\p)\xi({\mu^2\over 2},H_0+{\p^2\over
2}-E)\chi(\p)\cr
& & -\int
[dp_1dp_2dp_1'dp_2']
\chi^{\dag}(p_1+p_2)\bigg[\cr
& & \phi^{\dag}(p_1')\phi^{\dag}(p_2')
 {1\over H_0+\omega(p_1')+\omega(p_2')+\omega(p_1)+\omega(p_2)-E}
		\phi(p_1)\phi(p_2)\cr
& & +4(2\pi)^d\delta(p_1-p_1')\phi^{\dag}(p_2'){1\over
H_0+\omega(p_1')+\omega(p_2')+\omega(p_2)-E}\phi(p_2)
\bigg]\cr
& & \chi(p_1'+p_2').
\eeqs
Notice that the dependence on the cut-off of the previous operator is
traded for  a dependence on the renormalization scale \m{\mu}.

The resolvent
	operator  has the finite form
\beq
	R(E)={1\over H_0-E}+{1\over H_0-E}b^{\dag}{1\over
\Phi(\mu,E)}b{1\over H_0-E}
\eeq
which is the analogue of the Krein formula in the case of the manybody
problem.

\subsection{The Principal Operator}

Now let us understand the spectrum of the theory in the special case
 \m{d=2}. We will be mostly
interested in eignestates with \m{E<0}: the bound states of the
system. In each sector with a fixed number of particles we expect the
ground state to be of this form. 
( We will make some remarks later on the case \m{d=3} where we will
 see that the ideas in this section will not work  in that case.)

Then,
\beq
\xi({\mu^2\over 2},{\nu^2\over 2})={1\over 4\pi}\ln{\nu^2\over \mu^2}.
\eeq

We can rescale all the
momenta by \m{\surd(|E|)} to get
\beq
	\Phi(\mu,E)=\bigg[{1\over 2\pi}\ln{|E|\over \mu^2}+\;W\bigg],
\eeq
where
\beqs
	W&=&{1\over 2\pi}\int [d\p]\chi^{\dag}(\p)
\log\bigg[H_0+\omega(p)+1\bigg]\chi(\p)\cr
& & -\int
[dp_1dp_2dp_1'dp_2']
\chi^{\dag}(p_1+p_2)\bigg[\cr
& & \phi^{\dag}(p_1')\phi^{\dag}(p_2')
 {1\over H_0+\omega(p_1')+\omega(p_2')+\omega(p_1)+\omega(p_2)+1}
		\phi(p_1)\phi(p_2)\cr
& & +4(2\pi)^d\delta(p_1-p_1')\phi^{\dag}(p_2'){1\over
H_0+\omega(p_1')+\omega(p_2')+\omega(p_2)+1}\phi(p_2)
\bigg]\cr
& & \chi(p_1'+p_2').
\eeqs

Thus finding the bound state energy of our manybody system amounts to
finding the eigenvalues of \m{W}:
\beq
	W|\psi>=w\psi,\quad  E=-{\mu^2}e^{-2\pi w}
\eeq

The special case of the three body problem was studied in previous
papers \cite{hendersonrajeev}. We showed not only
that the ground state energy is finite, but that it can be estimated
by a simple variational ansatz.  Since we have described it elsewhere,
we won't elaborate on this point here.

\subsection{Bosonic Condensation in Two Dimensions}

We will now discuss the case of a large number of particles:
how to do transfinite quantum many body theory. We will only discuss
the mean field approximation, which should be good in the limit of a
large number of particles. In the ground state, we should expect all the
bosons to condense to a common state \m{u(p)}. Thus the problem is to
determine the pair of functions \m{u(p),\psi(p)}, where \m{\psi(p)}
being the wavefunction of the angel. The wavefunction \m{u(p)}
describes a condensate of bosons, which breaks translation
invariance. Of course there is no Bose-Einstein condensation for {\it
free} bosons in two dimensions. What we will show is that with an
attractive interaction of zero range there is in fact such a condensation.

The expectation value of the
operator \m{W} in the state \m{|u>,\psi>}  becomes, 
(for large \m{n}) the `principal
function' \m{U},
\beqs 	
 U&=&{1\over 2\pi}\int [dp] |\psi(p)|^2\log[n h_0(u)+\omega(p)+1]\cr
  & & -\int[dp_1dp_2dp_1'dp_2']\psi(p_1+p_2)^*\psi(p_1'+p_2')\cr
  & & \bigg[{u^*(p_1')u^*(p_2')u(p_1)u(p_2)n(n-1)
         \over
     nh_0(u)+\omega(p_1')+\omega(p_2')+\omega(p_1)+\omega(p_2)+1}\cr
 & & +4{(2\pi)^2\delta(p_1-p_1')nu^*(p_2')u(p_2)
      \over nh_0(u)+\omega(p_1')+\omega(p_2')+\omega(p_2)+1}\bigg]
\eeqs
where 
\beq
	h_0(u)=\int |u(p)|^2\omega(p)[dp].
\eeq
This \m{U} is to be minimized subject to the normalization conditions
\beq
	\int |u(p)|^2[dp]=\int |\psi(p)|^2[dp]=1.
\eeq
To get more explicit answers, let us ignore all except the leading
terms as \m{n\to \infty}:
\beq
	U=-n{\bigg|\int
\psi^*(p_1+p_2)u(p_1)u(p_2)[dp_1dp_2]\bigg|^2\over h_0(u)}+{1\over
2\pi}\log n+{\rm O}({ n^0}).
\eeq
Thus the ground state energy \m{E_n} of such a system of \m{n}
particles is given by 
\beq
	E_n=-\mu^2{e^{2\pi n \over \xi}\over n}\big[C_1+O({1\over
n})\big]
\eeq
where
\beq
	\xi=\inf_{u,\psi}{
\int|\psi(p)|^2[dp]\int|u(p)|^2[dp]\int\omega(p)|u(p)|^2[dp]
\over\bigg|\int
\psi^*(p_1+p_2)u(p_1)u(p_2)[dp_1dp_2]\bigg|^2 }.
\eeq
We will determine \m{\xi} by solving the variational
problem. Determination of the \m{C_1} takes more work which we will not
carry out here.
:  it is not  needed to determine the leading large \m{n} behavior

It is more convenient to use  the corresponding expressions in
position space,  with 
\m{\tilde u(x)=\int u(p)e^{ip\cdot x}[dp]} etc.
\beq
	\xi=\inf_{\tilde u,\tilde\psi}{
\int|\tilde\psi(x)|^2dx \int|\tilde u(x)|^2dx\int\half|\nabla \tilde u(x)|^2dx
\over\bigg|\int
\tilde\psi^*(x)\tilde u^2(x)dx\bigg|^2 }.
\eeq 
Eliminating \m{\tilde\psi} gives
\beq
	\xi=
\inf_{\tilde u}I[u]
\eeq
where,
\beq
I[\tilde u]={\int|\tilde u(x)|^2dx\int\half|\nabla \tilde u(x)|^2dx
\over \int
\tilde|\tilde u(x)|^4 dx }.
\eeq 
We can see that this amounts to solving a nonlinear differential
equation, which can be derived  by minimizing \m{\log I[u]}:
\beq
\nabla^2\tilde u -\beta \tilde u +g |\tilde u|^2 \tilde u=0.
\eeq
Here,
\beq
	\beta={\int |\grad \tilde u|^2 d^2x \over \int |\tilde u|^2 d^2x},\quad 
g={\int |\grad \tilde u|^2 d^2x \over \int |\tilde u|^4 d^2x}.
\eeq

The above partial differential equation (but without the constraints
on \m{\beta} and \m{g}) has been studied in
\cite{berger}, Theorem 6.7.25. ( We just need the special case of
dimension two.)  There is a normalizable solution for
every positive \m{\beta} and \m{g}. In fact it is enough to find the
solution for a particular pair of values of these constants. For
example, let \m{\tilde u_1} be a solution of 
\beq
\nabla^2\tilde u_1 - \tilde u_1 + |\tilde u_1|^2 \tilde u_1=0.
\eeq	 
Then, \m{u(x)=au_1(bx)} will solve the general equation with 
\beq
	\beta=b^{-2},\quad g=a^2b^{-1}.
\eeq
The quantity \m{I[u]} is invariant under such scale transformations:
 \m{I[u_1]=I[u]}. So it is enough to solve the special case.

Being the
ground state of a  many body problem, we should expect the solution to
be real  and have
circular symmetry around a point . Thus we only  have to solve an ODE, a
kind of nonlinear Bessel's equation
\beq
	v''(r)+{1\over 4r^2}v(r)+{v^3(r)\over r}=v(r)
\eeq
with \m{\tilde u(x)=\surd r v(r)} and \m{r=|x|}. That a square integrable 
 solution exists is proven\footnote{There is a  typo in
equation \m{\dag} of ref. \cite{berger}, p. 384; the \m{{1\over 4 r^2}
v(r)}  term is missing. It does not appear to affect the rest of the argument.}
 in ref. \cite{berger}.

We can  get a value for
\m{\xi\sim 12}  by numerical solution of the ODE. We  also
get this way the shape of the `soliton' (or `condensate' ); i.e., 
the wavefunction of  the bound state of a large number of bosons with
pointlike interactions. We plot below the solution \m{u(r)}; as shown
above, the
scales of the \m{u} and \m{r} axes are not physically relevant. The
wavefunction is peaked at the origin and decays exponentially at
infinity. The ground state thus breaks translation invariance but not
rotation invariance.\footnote{ I thank Govind Krishnaswami for solving
the ODE numerically and producing the graph.}

\includegraphics{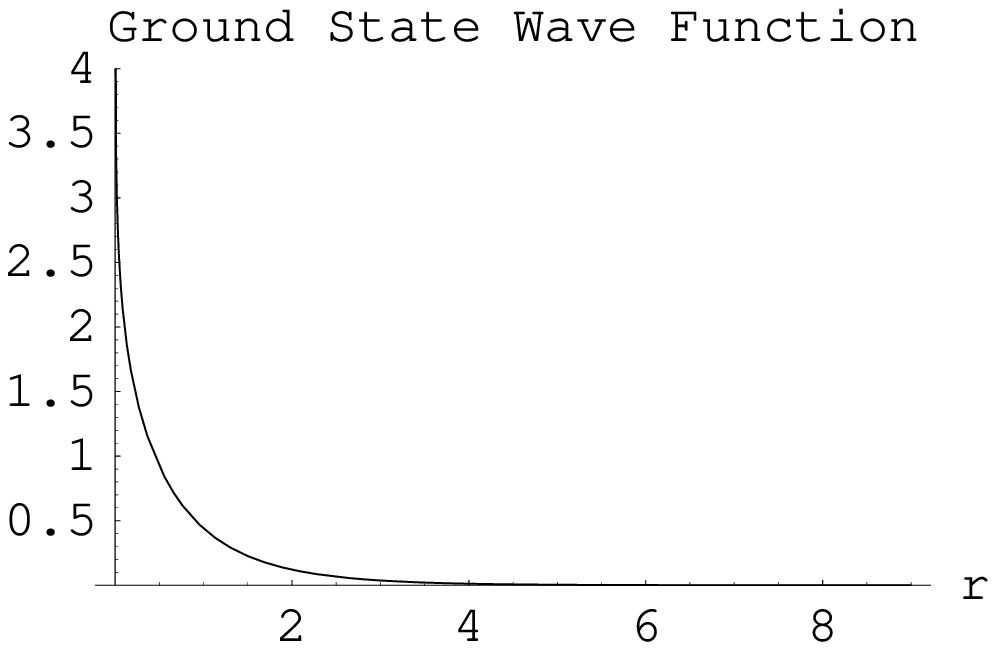}

One important conclusion of our analysis is that the magnitude of the 
 ground state energy grows exponentially with the
number of particles:
\beq
	E_n\sim-\mu^2e^{{\pi\over 6 }n}.
\eeq 
If we had a non-singular pair--wise interaction, we would expect the
 magnitude of the  ground state energy to grow like the number of
 pairs; i.e., like \m{n^2}.  

\subsection{Vortices in Superconductors}

A limiting case of the Landau-Ginzberg theory of superconductivity
gives  a realization of the example we have been discussing. The
lagrangian of this field theory is 
\beq
	L=\int |\nabla\phi|^2 d^3x+\half\int |{\rm curl} A|^2
d^3x+\int \lambda(|\phi|^2-a^2)^2.
\eeq
Here, \m{\phi} is a complex valued function on \m{R^3}, \m{A} a
covariant vector field describing the gauge potential, and
\beq
	\nabla \phi=\pdr \phi-ieA\phi
\eeq
the covariant derivative. If \m{\lambda>1} (Type II superconductor)
 this Lagrangian has a  static solution that is cylindrically
symmetric around a point in a plane  \m{R^2} and translationally invariant in
the orthogonal direction, carrying a unit of magnetic flux. Alontthe
axis of symmetry, the field \m{\phi} vanishes. Two such
vortices have a repulsive interaction: the energy decreases as the
position of the zeros of \m{\phi} move further apart in the plane. (
For given flux per unit area,
the solution that minimizes energy density is a triangular lattice-the
Abrikosov lattice.) As \m{\lambda\to 1^+} the force vanishes and the
energy is independent of the position ofthe zeros. (As \m{\lambda\to
1^-},
the force vanishes at finite distances, but  leads to  a
residual attractive contact interaction.) In fact for any
integer \m{n} there is a solution carrying \m{n} units of magnetic
flux, with  zeros of \m{\phi} at any prescribed set of \m{n} points on
the  plane \cite{taubes}.

Thus this vortices behave like identical free particles, located at
the position of the zeros of the field \m{\phi}. We can now consider
an approximate quantization of this system where onoy the modes of the
field that carry  infinitesimal energy are excited. This is just the
quantum mechanics of \m{n} identical particles on the plane
( The classical thermodynamics of this system 
has been studied in \cite{manton}.) In the
simplest quantization scheme,  we can assume these particles are
bosons.  There is no potential energy between them, if \m{\lambda} has
exactly the critical value.

However, as we saw earlier, there might be more subtle interactions
that arise from the boundary conditions on the wavefunctions; the
boundary being the region where a pair of particles come
together. Such contact interactions seem to have been ignored in the
literature on this subject:they arise if we approach the limit
\m{\lambda=1} from below rather than above. 
These lead to bound states of vortices,
with an energy that is determined by a new short distance scale
\m{\mu} exactly as discussed previously. In fact it is also possible
to have a condensate of a large number \m{n} of vortices with a mean
density of vortices (or magnetic flux) given by the function
\m{|u(r)|^2} determined in
the last section. It is of much interest to serach experimentally
 for such a condensation
of vortices in supercondutors which are on the borderline between type
I and type II. The prediction that the energy of such a configuration
depends exponentially on the number of vortices also should be tested 
experimentally.

\section{ Scalar Field in  Three dimensions: a Cautionary Tale}

Now let us see why the above approach cannot work as it stands in
three dimensions. Although a renormalized resolvent and a principal
operator can be constructed, the spectrum is not bounded
below for more than two particles. (However there may be other
aproaches that give a sensible formulation of this problem.
 See e.g., \cite{lethomas}.

We get again, a Krein formula:
\beq
	R(E)={1\over H_0-E}+{1\over H_0-E}b^{\dag}{1\over\Phi(\mu,E)}b{1\over H_0-E}.
\eeq

The eigenvalues are given by \m{\Phi(E)|\psi>=0}. Again, we can rescale all the
momenta by \m{\surd|E|} to get
\beq
	\Phi(\mu,E)=\bigg[-{\mu\over 16\pi}+\surd|E|\;W\bigg],
\eeq
where
\beqs
	W&=&{1\over 8\pi}\int [d\p]\chi^{\dag}(\p)
\surd\bigg[H_0+\omega(p)+1\bigg]\chi(\p)\cr
& & -\int
[dp_1dp_2dp_1'dp_2']
\chi^{\dag}(p_1+p_2)\bigg[\cr
& & \phi^{\dag}(p_1')\phi^{\dag}(p_2')
 {1\over H_0+\omega(p_1')+\omega(p_2')+\omega(p_1)+\omega(p_2)+1}
		\phi(p_1)\phi(p_2)\cr
& & +4(2\pi)^d\delta(p_1-p_1')\phi^{\dag}(p_2'){1\over
H_0+\omega(p_1')+\omega(p_2')+\omega(p_2)+1}\phi(p_2)
\bigg]\cr
& & \chi(p_1'+p_2').
\eeqs

This time  the new eigenvalues of the interacting system are given in terms
of the eigenvalues \m{w_n} of \m{W} by
\beq
	E_n=-\bigg({\mu\over 16\pi}\bigg)^2{1\over w_n^2}.
\eeq
The operator \m{W} itself depends on no  parameters: it determines the spectrum
of the interacting theory in terms of the parameter \m{\mu}. Our
system will have a finite ground state energy if and only if \m{W} is
strictly positive. The  eigenvalue closest to zero of \m{W} will give the
ground state energy by the above formula.

So far everything looks fine: it looks like the same approach as
in two dimensions is going to work. But we will now see that in the
three body sector the operator \m{W} has a zero eigenvalue which mens
the energy of the ground state is infinite.

In the sector of interest we have just one angel
and one boson. The states  will have the form
\beq
	|\psi>=\int
	[d\p][d\k]\psi(\p|\k)\phi^{\dag}(\k)\chi^{\dag}(\p)|0>.
\eeq
Only the first two terms in the Principal Operator will  constribute
to this sector. The eigenvalue problem for \m{W} becomes an integral
equation
for \m{\psi}:
\beq
{1\over 8\pi}\surd({\p^2\over 2}+\k^2+1)\psi(\p|\k)-
\int [d\k']{\psi(\p+\k-\k'|\k')\over
1+\k^2+2\k'^2-2\p\cdot\k'+\p^2}=
{\mu\over 16\pi\surd|E|}\psi(\p|\k)
\eeq
The total momentum \m{\p+\k} of the angel and the boson is
conserved. By going to the center of mass frame, we impose \m{\p=-\k}:
\beq
[1+{3\over 2}\k^2]u(\k)-8\pi\int[d\k']{u(\k')\over \half
+\k^2+\k'^2+\k\cdot \k'}={\mu\over \surd|E|}u(\k)
\eeq
Moreover it is
reasonable to expect that the ground state will have  zero angular
momentum: \m{u(\k)=u(k)}.  Performing the angular integration gives
\beqs
4\pi\int[d\k']{u(\k')\over \half
+\k^2+\k'^2+\k\cdot \k'}&=&
4\pi {1\over (2\pi)^d}(2\pi)\int_0^\infty k'^2u(k')dk'\cr
& & \int_{-1}^1d(\cos \theta)
{1\over \half +k^2+k'^2+k k'\cos\theta}\cr
&=&{1\over \pi}\int_0^\infty {k'u(k')\over k} dk'
\ln\bigg[{\half+ k^2+k'^2+ k k'\over \half+ k^2+k'^2- k k'}\bigg].\nonumber
\eeqs
Then, putting
\beq
	v(k)=ku(k)
\eeq
we get
\beq
	\bigg[1+{3\over 2} x^2\bigg]^{\half}v(x)-
{2\over \pi}\int_0^\infty v(y)\
\log\bigg[{\half+x^2+y^2+xy\over \half+x^2+y^2-xy}\bigg]dy={\mu\over
\surd|E|}v(x).
\eeq

We will now ask whether  the ground state energy determined by this
integral equation is finite; in other words  if there are
solutions to this equation for large \m{|E|}. We use  the same
ideas as the corresponding argument in two dimensions and see why they
break down
\cite{bruch},\cite{hendersonrajeev}.

We rewrite the equation as
\beq
v(x)=\int_0^\infty U_E(x,y)v(y)dy
\eeq
where
\beq
	U_E(x,y)={1\over \pi\bigg[-{\mu\over
		\surd|E|}+\surd\big(1+{3\over 2}x^2\big)\bigg]}
\log\bigg[{\half+x^2+y^2+xy\over \half+x^2+y^2-xy}\bigg]
\eeq
The question is whether this equation has a normalizable solution for
large \m{|E|}; if such a solutions exists, the three body problem has a
divergent ground state energy even after our renormalization.

The idea in two dimension is to   show that the Hilbert-Schmidt norm
\m{||U_E||_2} of this
integral kernel is less than one as \m{|E|} becomes large . Then the   equation
\m{v=K_Ev} will
have no   solution for large \m{|E|}. But in our case this will
diverge.   In fact,
\beq
||K_E||_2^2=\int_0^\infty\int_0^\infty K_E(x,y)^2dxdy
\eeq
Consider first the integral over \m{y}:
\beq
g(x)=\int_0^\infty
\bigg\{\log\bigg[{\half+x^2+y^2+xy\over \half+x^2+y^2-xy}\bigg]
 \bigg\}^2dy.
\eeq
We will first show that there exists a constant \m{C} such that
\beq
	g(x)\to  Cx
\eeq
for large  \m{x}. By scaling \m{y\to xy} this equivalent to
showing that
\beq
	\int_0^\infty
\bigg\{\log\bigg[{{1\over 2x^2}+1+y^2+y\over
{1\over 2x^2}+1+y^2-y}\bigg]
\bigg\}^2dy\to  C
\eeq
for large  positive \m{x}.
Now,
\beq
	\int_0^\infty
\bigg\{\log\bigg[{{1\over 2x^2}+1+y^2+y\over
{1\over 2x^2}+1+y^2-y}\bigg]
\bigg\}^2dy\to \int_0^\infty
\bigg\{\log\bigg[{1+y^2+y\over 1+y^2-y}\bigg]
\bigg\}^2dy=C
\eeq
The integral converges (in fact has a value of about \m{C\sim3.489}). Thus we
see that
\beqs
	||K_E||^2&=&\int_0^\infty{1\over \pi^2\bigg[-{\mu\over
		\surd|E|}+\surd\big(1+{3\over
2}x^2\big)\bigg]^2}\;\;g(x)dx
\eeqs
But this diverges logarithmically, since the integrand goes like
\m{1\over x} for large \m{x}. The corresponding integral in two
 dimensions converges  which makes the problem well-posed there.  This
is why the proof breaks down here.

In fact we can just put \m{|E|=\infty} and see that there is a
normalizable solution for \m{v}. It can be determined by iterating the
integral equation starting from the
trial function \m{v_0(x)=(1+{3\over 2}x^2)^{-1}}. We spare the reader
the details.

\section{Appendix: An Elementary  Formula for Inverses}

Suppose a self-adjoint  operator \m{X:\hi\to\hi} can be split into \m{2\times
2} blocks,
\beq
	X=\pmatrix{a&b^{\dag}\cr b&d\cr}
\eeq
with respect to a splitting \m{\hi=\hi_1\oplus\hi_2}
\beq
	a:\hi_1\to\hi_1,\quad  b^{\dag}:\hi_1\to\hi_2,\quad
b:\hi_2\to\hi_1,\quad d:\hi_2\to\hi_2.
\eeq
Let the  inverse be split similarly:
\beq
	X^{-1}=\pmatrix{\alpha&\beta^{\dag}\cr \beta&\delta\cr}.
\eeq
Then,
\beq
	a\beta^{\dag}+b^{\dag}\delta=0,\quad b\alpha+d\beta=0
\eeq
and
\beq
	a\alpha +b^{\dag}\beta=1,\quad b\beta^{\dag}+d\delta=1.
\eeq
Solving the first pair of equations we get two expressions for
\m{\beta}:
\beq
	-\delta b a^{-1}=\beta=-d^{-1}b\alpha.
\eeq
Putting either of these into the equation for \m{\alpha} gives two
expressions for it:
\beq
	a^{-1}+a^{-1}b^{\dag}\delta b a^{-1}=\alpha
	         =[a-b^{\dag}d^{-1}b]^{-1}.
\eeq
Similarly,
\beq
	d^{-1}+d^{-1}b \alpha  b^{\dag} d^{-1}=\delta
	      =[d-ba^{-1}b^{\dag}]^{-1}
\eeq
By combining the two ways of writing $\alpha$ and \m{\delta} we get
\beq
	\alpha=a^{-1}+a^{-1}b^{\dag}[d-ba^{-1}b^{\dag}]^{-1}b  a^{-1},
\eeq
\beq
	\delta=d^{-1}+d^{-1}b[a-b^{\dag}d^{-1}b]^{-1}b^{\dag}  d^{-1}.
\eeq
These identities will be used in the text.

\end{document}